\documentstyle[psfig]{mn}

\newcommand{\bc}{\begin{center}}
\newcommand{\be}{\begin{equation}}
\newcommand{\ee}{\end{equation}}
\newcommand{\ec}{\end{center}}

\newcommand{\ltsima}{\mbox{$\; \buildrel < \over \sim \;$}}
\def \simlt{\lower.5ex\hbox{\ltsima}}            
\def \gtsima{\mbox{$\; \buildrel > \over \sim \;$}}
\def \simgt{\lower.5ex\hbox{\gtsima}}            

\newcommand{\x}{${\bf x}$}

\newcommand{\q}{${\bf q}$}
\renewcommand{\S}{${\bf S}$}

\newcommand{\LCDMd}{$\Lambda$CDM}
\newcommand{\LCDMt}{$\Lambda$CDM128}
\newcommand{\msun}{$M_\odot$}
\newcommand{\fmax}{$F_{\rm max}$}
\newcommand{\rmax}{$R_{\rm max}$}
\newcommand{\vmax}{${\bf v}_{\rm max}$}
\newcommand{\fcl}{$f_{\rm cl}$}
\newcommand{\funcl}{$f_{\rm split}$}
\newcommand{\fov}{$f_{\rm ov}$}
\newcommand{\rlag}{$R_{\rm N}$}
\newcommand{\Ssim}{${\bf S}^{\rm sim}$}
\title[PINOCCHIO]
{PINOCCHIO: pinpointing orbit-crossing collapsed hierarchical
objects in a linear density field.}
\author[Monaco, Theuns \& Taffoni]{Pierluigi Monaco$^1$, Tom Theuns$^2$ and
Giuliano Taffoni$^3$\\
$^1$Dipartimento di Astronomia, Universit\`a di Trieste, 
via Tiepolo 11, 34131 Trieste, Italy\\
$^2$Institute of Astronomy, Madingley Road, Cambridge CB3 0HA, UK\\
$^3$SISSA, via Beirut 4, 34013 Trieste, Italy}

\begin{document}
\maketitle

\begin{abstract}
  PINOCCHIO (PINpointing Orbit-Crossing Collapsed Hierarchical
  Objects) is a new algorithm for identifying dark matter halos in a
  given numerical realisation of the linear density field in a
  hierarchical universe (Monaco et al. 2001).  Mass elements are
  assumed to have collapsed after undergoing orbit crossing, as
  computed using perturbation theory.  It is shown that Lagrangian
  perturbation theory, and in particular its ellipsoidal truncation,
  is able to predict accurately the collapse, in the orbit-crossing
  sense, of generic mass elements.  Collapsed points are grouped into
  halos using an algorithm that mimics the hierarchical growth of
  structure through accretion and mergers.  Some points that have
  undergone orbit crossing are assigned to the network of filaments
  and sheets that connects the halos; it is demonstrated that this
  network resembles closely that found in N-body simulations.  The
  code generates a catalogue of dark matter halos with known mass,
  position, velocity, merging history and angular momentum.  It is
  shown that the predictions of the code are very accurate when
  compared with the results of large $N$-body simulations that cover a
  range of cosmological models, box sizes and numerical resolutions.
  The mass function is recovered with an accuracy of better than 10
  per cent in number density for halos with at least $30-50$
  particles.  A similar accuracy is reached in the estimate of the
  correlation length $r_0$.  The good agreement is still valid on the
  object-by-object level, with 70-100 per cent of the objects with
  more than 50 particles in the simulations also identified by our
  algorithm.  For these objects the masses are recovered with an error
  of 20-40 per cent, and positions and velocities with a root mean
  square error of $\sim$1-2 Mpc (0.5-2 grid lengths) and $\sim$100
  km/s, respectively.  The recovery of the angular momentum of halos
  is considerably noisier and accuracy at the statistical level is
  achieved only by introducing free parameters.  The algorithm
  requires negligible computer time as compared with performing a
  numerical $N$-body simulation.
\end{abstract}

\begin{keywords}
cosmology: theory -- galaxies: halos -- galaxies: formation --
galaxies: clustering -- dark matter
\end{keywords}

\section{Introduction}
In Dark Matter (DM) dominated cosmological models, structure grows
through the gravitational amplification and collapse of small
primordial perturbations, imprinted at very early times by some
mechanism such as inflation. In particular in the case of Cold Dark
Matter (CDM), the formation of structure follows a hierarchical
pattern, with more massive halos forming from accretion of mass and
mergers of smaller objects (see e.g. Padmanabhan 1993 for a general
introduction). Galaxies form following the collapse of gas into these
dark matter potential wells (see, e.g., White 1996 for a review). An
accurate description of the non-linear evolution of perturbations in
the DM field is thus important for modeling the formation and evolution
of astrophysical objects within a cosmological setting.

The gravitational formation of dark matter halos is usually addressed
by means of $N$-body simulations.  However, a number of analytic or
semi-analytic techniques based on Eulerian or Lagrangian perturbation
theory (Bouchet 1997; Buchert 1997), for example the Press \&
Schechter (1974, PS) and similar techniques (see e.g. Monaco 1998 for
a recent review), were devised to approximate some aspects of the
gravitational problem.  Analytic techniques have the advantage of
being both fast and flexible, thereby giving insight into the dynamics
of the gravitational collapse.  In particular, Lagrangian Perturbation
Theory (LPT; Moutarde et al. 1992; Buchert \& Ehlers 1993; Catelan
1995) and more specifically its linear term, the Zel'dovich (1970)
approximation, were used to compute many properties of the density and
velocity fields in the \lq mildly non-linear regime\rq~ when the
density contrast is not very high, and particle trajectories still
retain some memory of the initial conditions.  The PS and extended PS
approaches (Peacock \& Heavens 1990; Bond et al. 1991; Lacey \& Cole
1993) were used to generate merger histories of DM halos. Extensions
of the PS approach to the non-linear regime were attempted by many
authors (Cavaliere, Colafrancesco \& Menci 1992; Monaco 1995, 1997a,b;
Cavaliere, Menci \& Tozzi 1996; Audit, Teyssier \& Alimi 1997; Lee \&
Shandarin 1999; Sheth \& Tormen 1999, 2001; Sheth, Mo \& Tormen
2001). Alternative approaches assumed objects to form at the peaks of
the linear density field (Peacock \& Heavens 1985; Bardeen et
al. 1986; Manrique \& Salvador-Sol\`e 1995; Bond \& Myers 1996a,b;
Hanami 1999), or applied the Zel'dovich approximation to smoothed
initial conditions (truncated Zel'dovich approximation, Coles, Melott
\& Shandarin 1993; Borgani, Coles \& Moscardini 1994), or used the
second-order LPT solution for the density field (Scoccimarro \& Sheth
2001), or joined linear-theory predictions with Monte-Carlo methods
such as the block-model (Cole \& Kaiser 1988) and merging-cell model
(Rodrigues \& Thomas 1996; Nagashima \& Gouda 1998; Lanzoni, Mamon \&
Guiderdoni 2000).

These approaches are limited to the linear or mildly non-linear regime
and are generally unable to recover accurately the wealth of
information available with a large numerical simulation. In
particular, although PS provides a reasonable first approximation to
the mass function of halos (Efstathiou et al. 1988; Lacey \& Cole
1994), it underestimates the number of massive objects and
overestimates the number of low mass ones (see, e.g., Gelb \&
Bertschinger 1994; Governato et al.  1999; Jenkins et al. 2001; Bode
et al. 2000). Similarly, the merger history of DM halos is reasonably
well reproduced by the extended PS approach, but there are systematic
differences when compared with simulations, and also some theoretical
inconsistencies (Lacey \& Cole 1993; Somerville \& Kolatt 1999; Sheth
\& Lemson 1999).  The clustering of halos of given mass in the PS
approach can be obtained analytically (Mo \& White 1996; Catelan et
al. 1998; Porciani, Catelan \& Lacey 1999; Sheth \& Tormen 1999;
Sheth et al. 2001; Colberg et al.  2001), but the extended PS approach
is not able to produce both spatial information and merger histories
at the same time.  This is true also for many of the non-linear
extensions of the PS approach mentioned above.  The merging cell model
can provide spatial information on the halos (Lanzoni et al. 2000),
but only in the space of initial conditions (Lagrangian space), while
the truncated Zel'dovich approximation, though able to predict
correlation functions in the Eulerian space, is not accurate in
predicting the masses of the single objects (Borgani et al. 1994).
Finally, the peak-patch approach (Bond \& Myers 1996a,b) can also
generate catalogues of halos with spatial information, but has never
been extended to predict the merging histories.

Semi-analytical models of galaxy formation assume that the properties
of a galaxy depend on the merger history of its associated DM halo. So
in order to make predictions of the clustering properties of galaxies
of a given type, one needs to be able to compute the merger history
and spatial clustering simultaneously. Given the limitations of the
analytic techniques discussed above, such models have usually resorted
to analysing large $N$-body simulations with very many snapshots to
reconstruct the merger histories (see. e.g., Diaferio et al. 1999).
Alternatively, the extended PS approach is used to compute the merger
histories, but $N$-body simulations to obtain the spatial information
on the halos statistically (e.g., Benson et al. 2000).

A new approach for obtaining the spatial information and the merger
history simultaneously for many halos was recently proposed by Monaco
et al.  (2001, hereafter paper I; see also Monaco 1999 for preliminary
results). In the PINOCCHIO (PINpointing Orbit-Crossing Collapsed
HIerarchical Objects) formalism, LPT is used in the context of the
extended PS approach, as in Monaco (1995; 1997a,b) and Monaco \&
Murante (1999), to provide predictions for the collapse of fluid
elements in a given numerical realisation of a linear density
field. Mass elements are assumed to have collapsed after undergoing
orbit crossing. Such points are then grouped into halos using an
algorithm that mimics the hierarchical growth of structure through
accretion and mergers. The Zel'dovich approximation is used to compute
the Eulerian positions of halos at a given time. Some points that have
undergone orbit crossing are assigned to the network of filaments and
sheets that connects the halos. Paper~I contained a preliminary
comparison to simulations, demonstrating that PINOCCHIO can accurately
reproduce many properties of the DM halos from a large $N$-body
simulations that started from the same initial density field. The good
agreement is not only for statistical quantities such as the mass or
the correlation function, but extends to the object-by-object
comparison. PINOCCHIO thus provides a significant improvement over the
extended PS approach, which is known to be approximately valid only in
a statistical sense (Bond et al. 1991; White 1996).

In this paper, the PINOCCHIO code is described in more detail,
focusing on some aspects that were neglected in paper I, in particular
the validity of orbit crossing as definition of collapse, the ability
to disentangle halos from the filament web, a complete description of
the free parameters involved in the model, its validity at galactic
scales, and its extension to predicting the angular momentum of DM
halos.  An accompanying paper (Taffoni, Monaco \& Theuns 2001) focuses
on the ability of PINOCCHIO to recover the merging histories of DM
halos. The paper is organised as follows.  Section 2 presents the
first step of PINOCCHIO, the prediction of collapse time for generic
mass elements.  This prediction is directly compared with the results
of two different N-body simulations.  Section 3 presents the second
step of PINOCCHIO, the fragmentation algorithm, with attention to the
ability of separating filaments from relaxed halos.  In section 4
PINOCCHIO is applied to the initial conditions of the two simulations
mentioned above.  The results of PINOCCHIO are compared with the
N-body ones in terms of statistical quantities (mass and correlation
functions), on a particle-by-particle basis (mass fields) and on an
object-by-object basis (mass, position and velocity).  In Section 5
PINOCCHIO is extended to predict the angular momentum of the DM halos.
Section 6 discusses the relation of PINOCCHIO to previous analytic and
semi-analytic approximations to the gravitational problem.  Section 7
gives the conclusions.

\section{Predicting the collapse time from orbit crossing}

\subsection{The definition of collapse}

Linear theory is unable to treat the later stages of gravitational
collapse, because the density grows at a constant rate and hence never
becomes very high.  Therefore, it is usually assumed that collapse takes
place when the density contrast $\delta\equiv
(\rho-\bar{\rho})/\bar{\rho}$ reaches values $\sim 1$ (here
$\bar{\rho}(t)$ is the density of the background cosmological model).
In the special case of a spherical top-hat perturbation, a singularity
(a region of infinite density) forms when the corresponding linear
extrapolation of the density contrast reaches a value $\delta_c \simeq
1.686$. It is usually argued that the formation of the singularity
corresponds to the formation of the corresponding DM halo.

When more general cases than the spherical model are considered, the
very definition of collapse becomes somewhat arbitrary. Both in LPT and
in the evolution of ellipsoidal perturbations (White \& Silk 1979;
Monaco 1995, 1997a; Bond \& Myers 1996), collapse takes place along the
three different directions defined by the eigen vectors of the
deformation tensor, at three different times (see below). Therefore,
several definitions of collapse have been proposed, one related to
first-axis collapse (Bertschinger \& Jain 1994; Monaco 1995; Kerscher,
Buchert \& Futamase 2000), and another related to third-axis collapse
(Bond \& Myers 1996; Audit et al. 1997; Lee \& Shandarin 1998; Sheth et
al. 2001).  The difference between these two definitions is discussed
in detail in Section 5.1.

In the Lagrangian picture of fluid
dynamics the Eulerian (comoving) position \x\ of a fluid element (or
equivalently of a mass particle) is related to the initial
(Lagrangian) position \q\ through the relation:

\be {\bf x}({\bf q},t) = {\bf q} + {\bf S}({\bf q},t), \label{eq:lag} \ee

\noindent
where ${\bf S}({\bf q},t)$ is the displacement field.  The
Euler-Poisson system of equations (see Padmanabhan 1993) can be recast
into an equivalent set of equations for \S\ (see, e.g., Catelan 1995).
LPT is a perturbative solution to that system of equations, whose
first-order term is the well-known Zel'dovich approximation
(Zel'dovich 1970; Buchert 1992):

\be S_a({\bf q},t) = -b(t)\, \varphi_{,a}({\bf q}). \label{eq:zel} \ee

\noindent
Here and below, commas denote differentiation with respect to \q,
$b(t)$ is the linear growing mode, and $\varphi({\bf q})$ is the
rescaled peculiar gravitational potential, which obeys the Poisson
equation:

\be \nabla^2 \varphi({\bf q}) = \delta({\bf q},t_i)/b(t_i) 
\equiv \delta_l({\bf q}), \label{eq:poisson} \ee

\noindent
where $t_i$ is an initial time at which linear theory holds.  The
quantity $\delta_l({\bf q})$ does not depend on the initial time. It is
called linear contrast, as it is equal to the linear extrapolation of
the density contrast to the time defined by $b(t_0)=1$, which can be
taken to be the present time (i.e., $b(z=0)=1$).

As the fluid element contains by construction a fixed but vanishingly
small mass, its density can be written as the inverse of the Jacobian
determinant of the transformation given in equation~\ref{eq:lag}:

\be 1+\delta({\bf q},t) = \det\left(x_{a,b}\right)^{-1} 
= \det\left(\delta^K_{ab} + S_{a,b} \right)^{-1}. \label{eq:jac} \ee

\noindent
(Here $\delta^K_{ab}$ is the Kronecker symbol).  When the Jacobian
determinant vanishes, the density formally goes to infinity. This
corresponds to the formation of a caustic, a process discussed in
detail by Shandarin \& Zel'dovich (1989). At this time, the
transformation ${\bf x}\rightarrow{\bf q}$ becomes multi-valued, and
particle trajectories undergo orbit crossing (OC).

Because the density becomes high at OC, we identify this moment as the
collapse time (Monaco 1995; 1997a). In this way, collapse is well
defined and easy to compute using LPT which remains valid up to that
point but breaks down afterwards. We note that this definition
corresponds to first-axis collapse as discussed at the beginning of
this Section.  This definition of collapse does not require the
introduction of any free parameters. However, a drawback of this
definition is that it does not guarantee that the mass element is
going to flow into a relaxed DM halo. Indeed, a fraction of particles
that undergo OC remain in low density filaments instead of collapsed
halos. 

The calculation of collapse times is presented in Monaco (1997a), to
which we refer for all details.  LPT converges in predicting the
collapse time of a generic fluid element, as long as not more than 50 per cent
of mass has collapsed.  First-order LPT, i.e. the Zel'dovich
approximation, is exact (up to OC) in the case of planar symmetry, but
in the spherical limit, relevant for the collapse of high peaks, it
overestimates the growing mode at collapse time by nearly a factor of
two (the value of $\delta_l$ for spherical collapse is 3, compared with
1.686), while second-order LPT is ill-behaved in under densities.  Thus
third-order LPT must be used to calculate the collapse time of generic
mass elements.

The Lagrangian perturbative series can be truncated so as to resemble
formally the collapse of an ellipsoid in an external shear field (see
also Bond \& Myers 1996a).  When the peculiar gravitational potential
(equation \ref{eq:poisson}) is expanded into a Taylor series around a
generic position (taken to be the origin of the \q\ frame), the first
term relevant for the deformation of the fluid element is the quadratic
one, $\varphi({\bf q}) \simeq \varphi_{,ab} q_a q_b/2$.  In the
principal frame of $\varphi_{,ab}$ this can be written as:

\be \varphi({\bf q}) = \frac{1}{2} (\lambda_1 q_1^2 + \lambda_2 q_2^2
+ \lambda_3 q_3^2), \label{eq:potell} \ee

\noindent 
where $\lambda_i$ are the three eigenvalues of $\varphi_{,ab}$. The
initial conditions for the ellipsoid semi axes $a_i$ at the initial time
$t_i$ are:

\be a_i = a(t_i)(1-b(t_i)\lambda_i), \label{eq:ell} \ee

\noindent
where $a(t)$ is the scale factor. Note that at the initial time the
ellipsoid is an infinitesimally perturbed sphere.  With these initial
conditions, the exact equations of ellipsoidal collapse can be
integrated numerically (Monaco 1995; Bond \& Myers 1996a). However, it
easier to solve exactly the third-order LPT equations in the
ellipsoidal case of equation~\ref{eq:potell}, as only first and second
derivatives of the peculiar potential are retained.  This LPT solution
gives a very good approximation to the numerical integration in all
cases with the exception of the spherical limit.  A small numerical
correction is sufficient to recover properly this limit; this is
described in Appendix B of Monaco (1997a). Apart from describing
accurately the collapse of an ellipsoid, this solution gives a general
approximation for the LPT evolution of a mass element under the action
of gravity.  This approximation, which will be denoted by ELL in the
following, is easy to implement as it requires only the computation of
the deformation tensor, while full third-order LPT requires the
solution of many Poisson equations, thereby introducing numerical
noise. Moreover, 3rd-order LPT still under predicts the quasi-spherical
collapse of the highest peaks (a simple correction, as in the ELL case,
is not feasible in this case), and consequently also the high-mass tail
of the mass function. In general, ELL is an adequate approximation to
compute the OC collapse time of generic mass elements.

In conclusion, it is worth stressing again that in this context ELL is
purely a convenient truncation of LPT; no constraint is put on the
shape of the collapsing objects, nor on the \lq shape\rq\ of the mass
elements (which is simply a meaningless concept).

\begin{table*}
\begin{tabular}{l|cccccccc}
\hline
  & $N_{\rm part}$ & $L_{\rm box}$ (Mpc/h) & $\Omega_0$ & $\Omega_\Lambda$
  & h & $\Gamma$ & $\sigma_8(z=0)$ & $M_{\rm part}$ (\msun) \\
\hline
SCDM    & 360$^3$ & 500 & 1.0 & 0.0 & 0.5  & 0.5   & 1.0 & $1.49\times 10^{12}$ \\
\LCDMd  & 256$^3$ & 100 & 0.3 & 0.7 & 0.65 & 0.195 & 0.9 & $7.64\times 10^{9}$ \\
\LCDMt  & 128$^3$ & 100 & 0.3 & 0.7 & 0.65 & 0.195 & 0.9 & $6.11\times 10^{10}$ \\
\hline
\end{tabular}
\caption{Simulations used for the analysis.}
\end{table*}

\subsection{Testing OC as definition of collapse}

Before using OC as collapse prediction, it is necessary to decide
whether LPT (and ELL in particular) is accurate enough to reproduce the
OC-collapsed regions, and how these are related to the relaxed halos.
This can be done by applying LPT to the initial conditions of a large
$N$-body simulation, and comparing the LPT OC regions to those computed
by the simulation.

For this and further comparisons we use two collision less
simulations.  The first, a standard CDM model (SCDM), has been
performed with the PKDGRAV code, and consists of 360$^3$
($\sim$46$\times$10$^6$) DM particles (Governato et al. 1999); it was
also used in paper I.  The second simulation has been performed with
the Hydra code (Couchman, Thomas \& Pearce 1995), and consists of
256$^3$ DM particles in a flat Universe with cosmological constant
(\LCDMd).  In order to test for resolution effects, the same
simulation has been run with 128$^3$ particles, resampling the initial
displacements on the coarser grid (we will refer to it as \LCDMt).
The main characteristics of the simulations are summarised in Table 1.
These simulations allow us to test PINOCCHIO for different
cosmologies, different resolutions, and different N-body codes,
reaching a range of at least 5 orders of magnitude in mass with good
statistics in terms of both numbers of halos and numbers of particles
per halo.  The PKDGRAV simulation samples a very large volume, making
it suitable for testing the high mass tail of the mass function.  The
Hydra simulation samples a much smaller volume but at higher
resolution, so we can test the power-law part of the mass function at
small masses.  Note that in all the simulations the particles are
initially placed on a regular cubic grid.  We have compared our
results with another $\Lambda$CDM simulation performed with PKDGRAV,
with the same box (in Mpc/h) and number of particles as the SCDM one.
The comparison confirms all the results given in this paper, and is
not presented here.

The predictions of collapse are performed as follows.  The linear
contrast $\delta_l$ is obtained from the initial displacements of the
simulation using the relation (see equations~\ref{eq:zel} and
\ref{eq:poisson}):

\be \S_{a,a}({\bf q},t_i) = - \delta_l({\bf q}) b(t_i). \label{eq:ic} \ee

\noindent
For the SCDM simulation the displacements are first resampled on a
$256^3$ grid for computational ease.  In this case, as well as
throughout the paper, differentiations are performed with Fast Fourier
Transforms (FFTs). This procedure allows one to recover the linear
contrast with minimum noise and no bias.  The linear contrast
$\delta_l$ is then FFT-transformed and smoothed on many scales $R$
with a Gaussian window function in Fourier space:

\be \tilde{W}(kR) = \exp \left( -k^2 R^2/2 \right). \label{eq:gausmo}
\ee

\noindent
The smoothing radii are equally spaced in $\log R$, except for the
smallest smoothing radius which is set to 0 in order to recover all
the variance at the grid scale. The largest smoothing radius is set
such that the variance of the linear density contrast $\sigma(R_{\rm
max})=1.686/6$, making the collapse of a halo at this smoothing scale
approximately a 6 $\sigma$ event. The smallest non-zero smoothing
radius is set to a third of $R_{\rm max}$. Because of the stability of
Gaussian smoothing, 25 smoothing radii in addition to $R=0$ give
adequate sampling for a $256^3$ realisation (we use 15+1 smoothing
radii for 128$^3$ grids).  For each smoothing radius $R$ the
deformation tensor, $\varphi_{a,b}({\bf q},R)$, is obtained in the
Fourier space from the FFT-transformed, smoothed linear density
contrast $\tilde{\delta}_l({\bf k};R)$ as $\tilde{\varphi}_{a,b}({\bf
k};R) = - k_a k_b/k^2 \ \tilde{\delta}_l({\bf k};R) $, and then
transformed back to real space, again with FFT.
Double
precision is required in this calculation to obtain sufficiently
accurate results.  The ELL collapse times are computed for each grid
point from the value of the deformation tensor as described in Section
2.1 and Appendix B of Monaco (1997a).

It is convenient to use the growing mode $b(t)$ as time variable,
because with this choice the dynamics of gravitational collapse is
(almost) independent of the background cosmology (see, e.g., Monaco
1998).  For this reason, in place of the collapse time $t_c$ we record
the growing mode at collapse, $b_c=b(t_c)$.  With the procedure outlined above, a
collapse time is computed for each grid vertex ${\bf q}$ and for each
smoothing radius $R$, i.e. $b_c=b_c({\bf q};R)$.  We define the
inverse collapse time field $F$ as:

\be F({\bf q}; R) \equiv 1/b_c({\bf q};R). \label{eq:fdef} \ee

\noindent
In the case of linear theory $F=\delta_l/\delta_c$.  The values of the
$F$-field at a single point ${\bf q}$ correspond to the trajectories
in the $F-R$ plane (or equivalently the $F-\sigma^2(R)$ plane) used in
the excursion set approach to compute the mass function. In fact, as
shown by Monaco (1997b), this quantity is obtained from the absorption
rate of the $F(R)$ trajectories by a barrier put at a level $F_c$.  As
the smoothing filter is Gaussian, these trajectories {\it are not}
random walks but are strongly correlated.  In general, the computation
of the absorption rate requires no free parameter as long as the
collapse condition does not.  To solve the cloud-in-cloud problem, we
record for each grid point the {\it largest} radius $R_c$ at which the
inverse collapse time overtakes $F_c$; the grid point is assumed to be
collapsed at all smaller scales.  We call this radius $R_c({\bf q})$,
the {\it collapse radius} field (Monaco \& Murante 1999).  $R_c$
depends on the height of the barrier as well as on time.

The $R_c$ field for the simulations is obtained as follows. The
displacement field \Ssim\ (i.e. the displacement of N-body particles
from their initial position on the grid) is smoothed in the Lagrangian
space \q\ with the same set of smoothing radii. (Also here, we
resample the large 360$^3$ simulation to a 256$^3$ grid using nearest
grid point interpolation.) Each smoothed field is differentiated using
FFTs along the three spatial directions and the Jacobian determinant
$\det(\delta^K_{ab} + S^{\rm sim}_{a,b})$ is computed for each grid
vertex. For each grid point, we again record the largest smoothing
radius $R_c^{\rm sim}({\bf q})$ at which the Jacobian determinant
first becomes negative (hence passing through 0).

\begin{figure*}
\centerline{
\psfig{figure=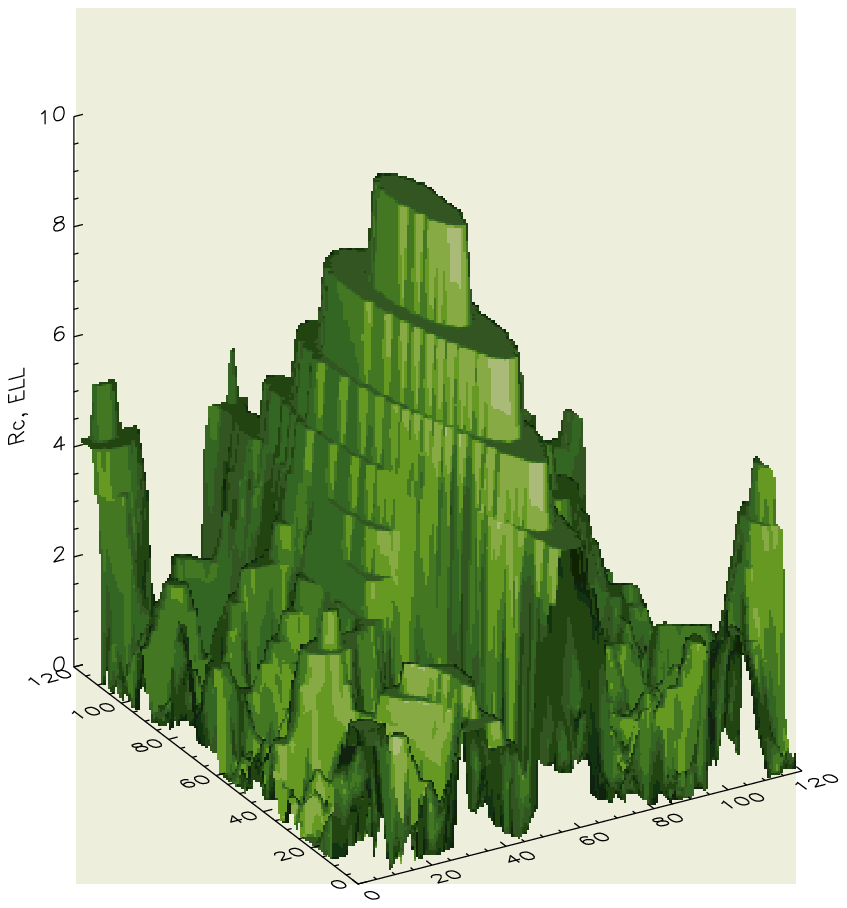,width=9cm}
\psfig{figure=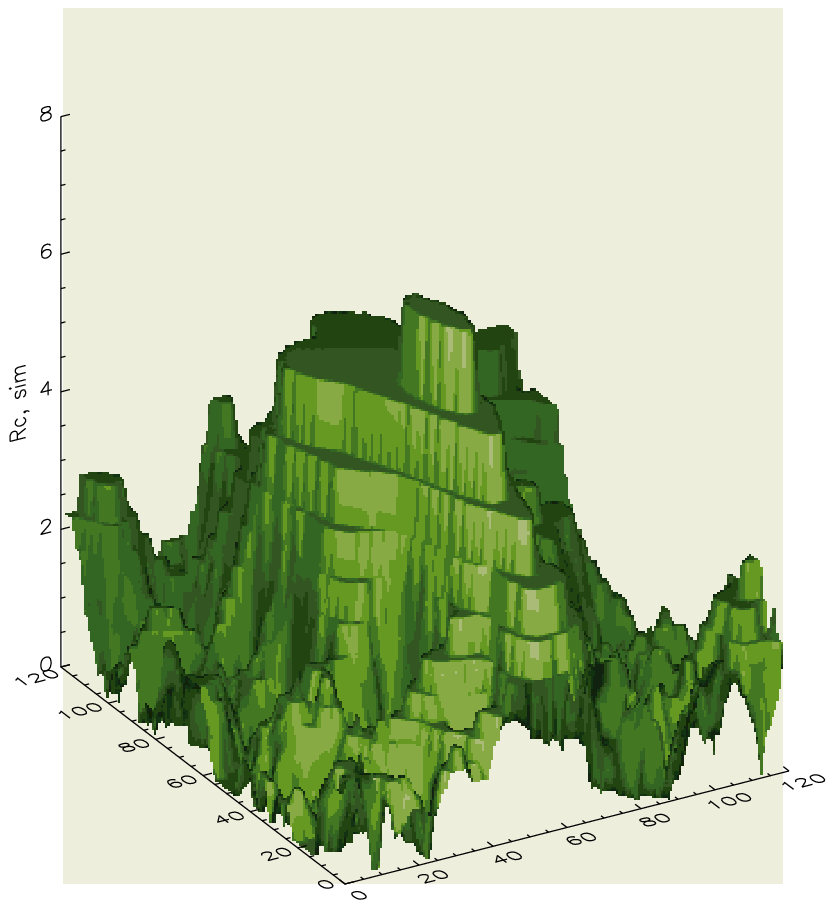,width=9cm}
}
\centerline{
\psfig{figure=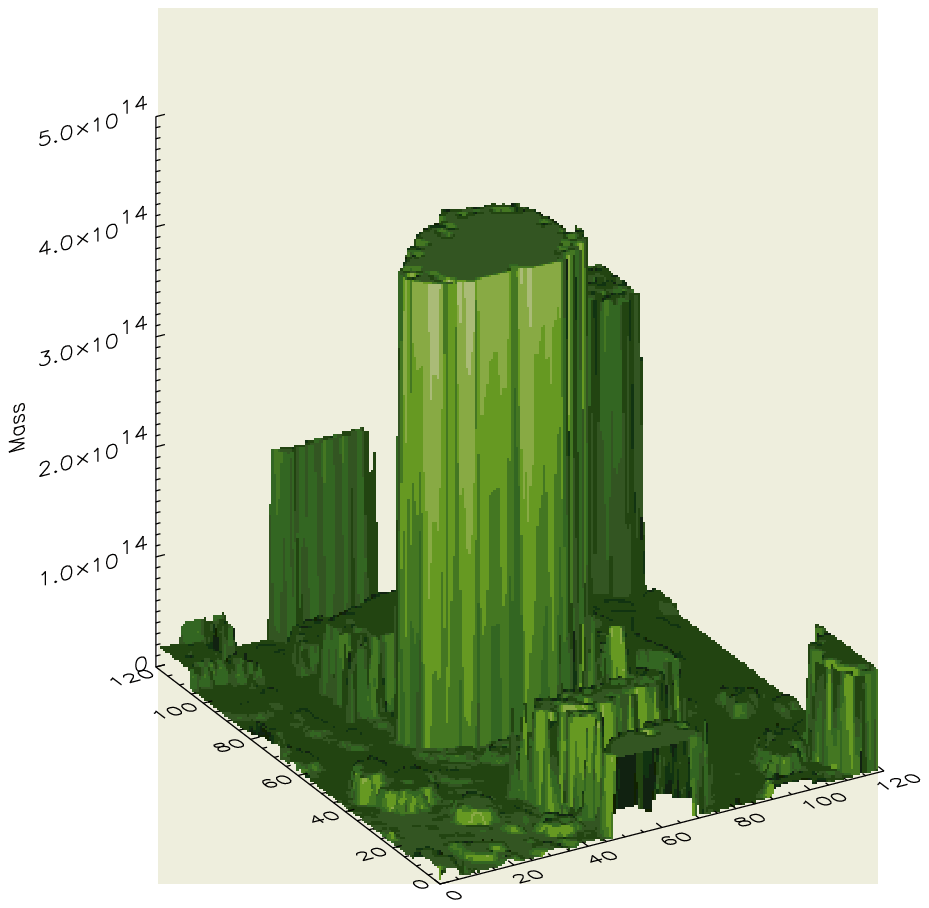,width=9cm}
\psfig{figure=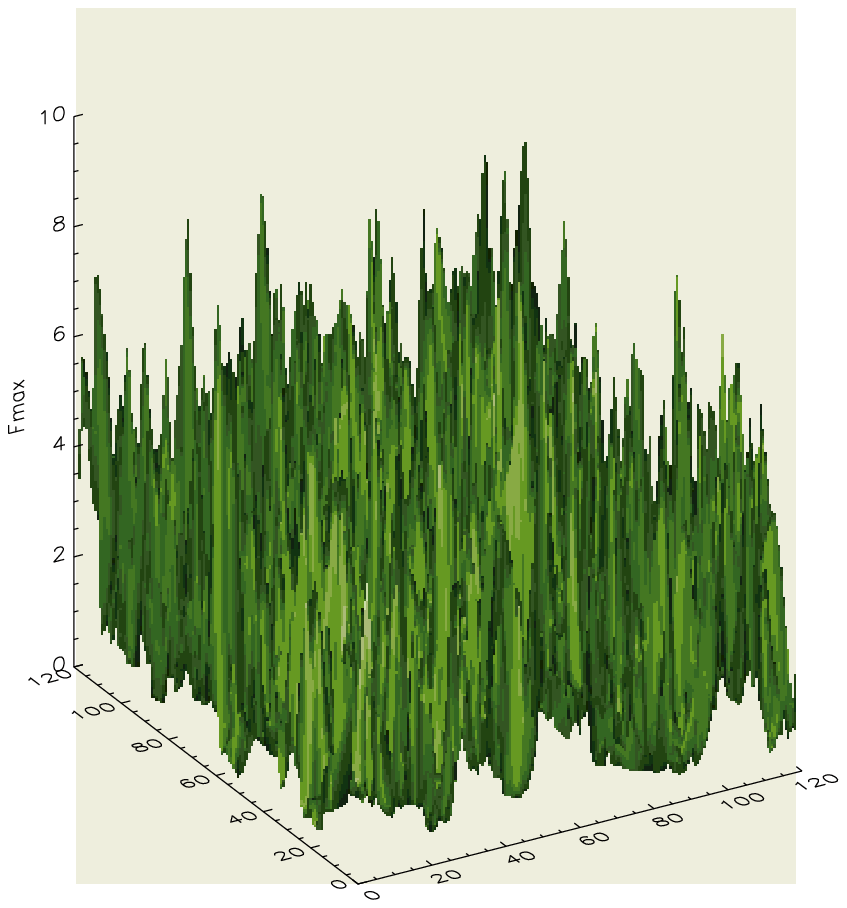,width=9cm}
}
\caption{Upper panels: collapse radius fields $R_c$ for a section of
the Lagrangian space of the \LCDMd\ simulation at redshift $z=0$. In
the left panel we show the ELL prediction, and in the right panel the
results from the simulation.  Lower left panel: mass field for the same
section; the mass field gives for each particle the mass of the halo it
belongs to at $z=0$.  Ungrouped particles are assigned 0 mass. Lower
right panel: inverse collapse time \fmax\ for the same section.}
\label{fig:fields}
\end{figure*}

\begin{figure*}
\centerline{
\psfig{figure=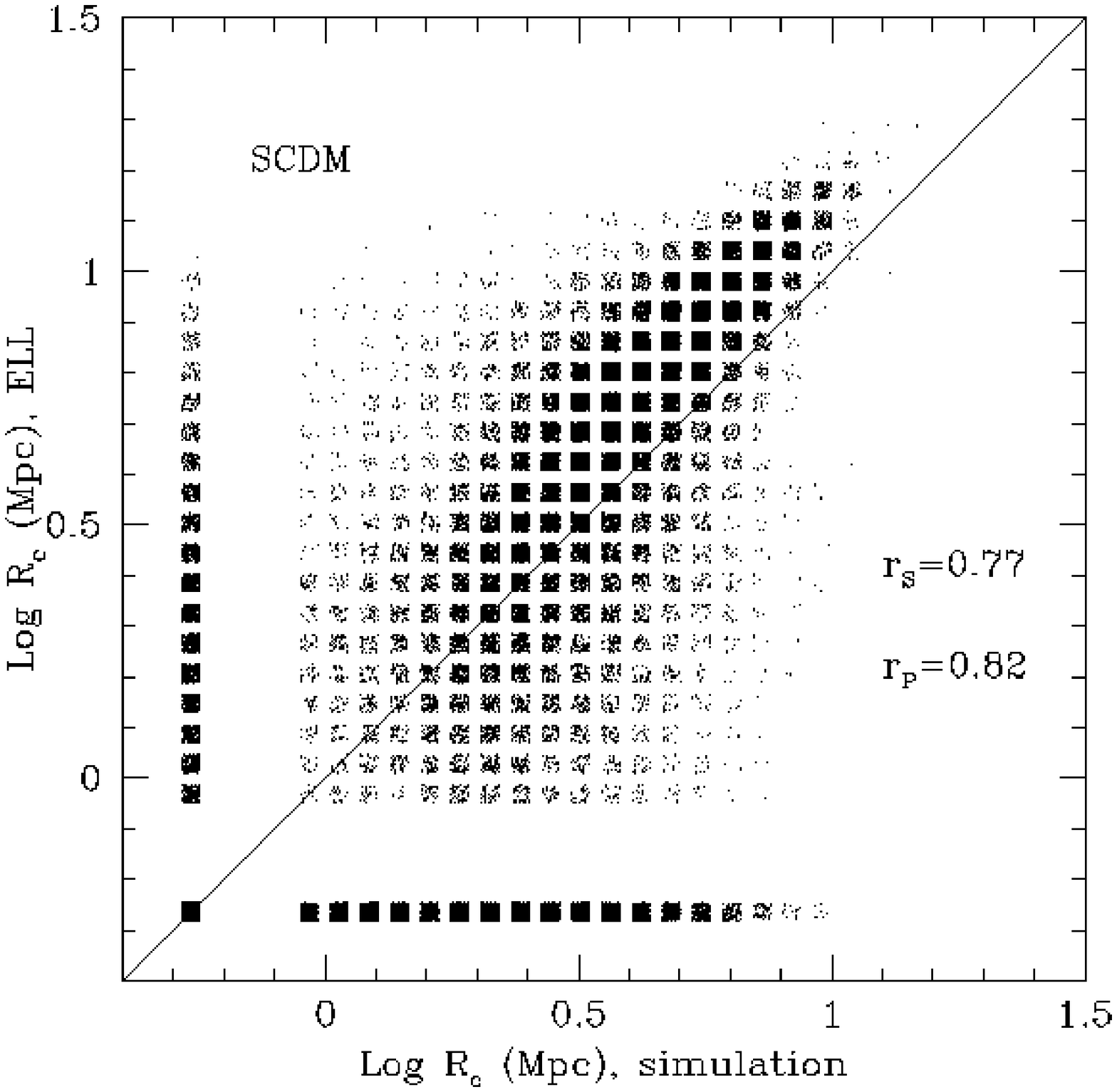,width=8cm}
\psfig{figure=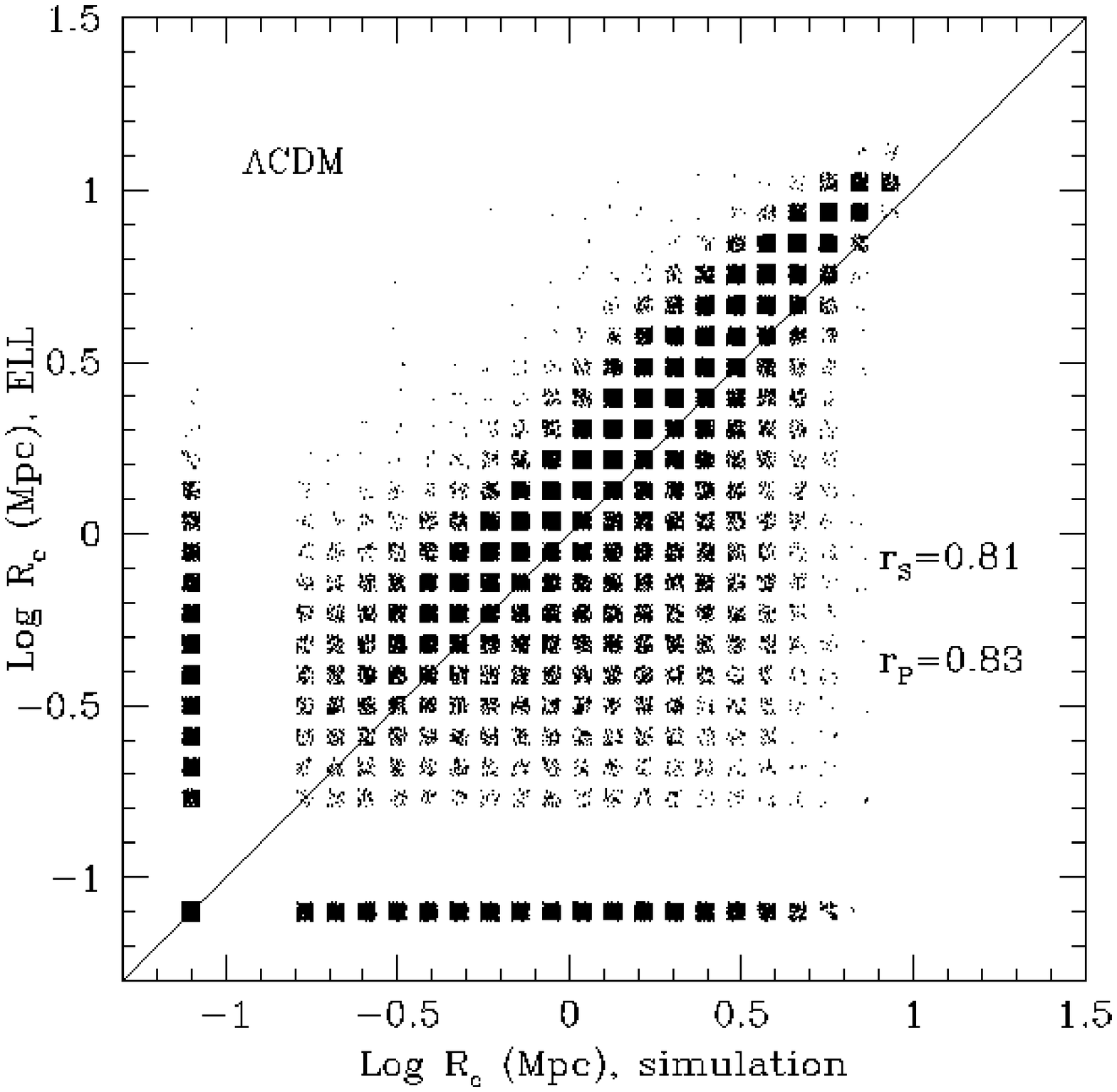,width=8cm}
}
\caption{Comparison of the collapse radius fields $R_c$, as predicted
by ELL, with the values found in the simulations, for a random sample
of $\sim$20000 points for the SCDM model (left panel), and the \LCDMd\
model (right panel).  For clarity some random noise has been added to
the discrete $R_c$ values, so that they lie on squares instead of
points.}
\label{fig:rc}
\end{figure*}

The $R_c$ field computed using LPT and obtained from the simulations at
redshift $z=0$ are compared in figure~\ref{fig:fields}. The two fields are
remarkably similar, exhibiting the same structure of broad peaks, with
the difference that the peaks of the simulation are lower, as
anticipated by Monaco (1999). In figure~\ref{fig:rc} we show a more
quantitative point-by-point comparison between the two fields. For
display purposes, some random noise has been added to the discrete
values of $R_c$; in this way the values lie in squares instead of
points. There is a reasonably tight correlation between the predicted
and numerical collapse-radius fields, which confirms the power of LPT
to predict the mildly non-linear evolution of perturbations; it is
noteworthy that this comparison does not involve free parameters.  The
correlation is quantified by the well-known Spearman rank correlation
coefficient $r_S$ and Pearson's linear correlation coefficient $r_P$,
both reported in the panels of figure~\ref{fig:fields}.  (A high value of
$r_S$ indicates the existence of a relation with moderate scatter, a
high value of $r_P$ indicates the existence of a good linear relation.)
The coefficients take rather high values of $\sim 0.8$, confirming in
an objective and quantitative way the correlation.

However, as noted also in figure~\ref{fig:fields}, the relation between
the two $R_c$ fields is not unbiased: the simulated $R_c$ field is
lower than the ELL one, especially at large $R$-values. The cause of
this behaviour can be understood as follows. LPT predicts that after
OC particles do not remain bound to the caustic region but move away
from it, in contrast to what happens in the simulations. Therefore, as
in this analysis particles are not explicitly restricted to the pre-OC
(single stream) regime, the displacements in the simulation are always
smaller than those predicted by LPT.  As a consequence, the collapse
radius obtained by the smoothed displacements of the simulation is
lower than that predicted by LPT.  This bias disappears at small
radii, which are however dominated by numerical noise.

The difference between the LPT and simulation fields $R_c$ can also be
quantified by the cumulative distribution of the $R_c$ fields as a
function of $R$, or equivalently of the variance $\sigma^2(R)$. We will
denote this function by $\Omega(<\sigma^2)$, since it is also the
fraction of mass collapsed on a scale $\ge R$ where the rms is smaller
than $\sigma$. This quantity is used in the PS approach to obtain the
mass function

\be M n(M) dM = \bar{\rho} \frac{d\Omega}{d\sigma^2} \left|
\frac{d\sigma^2}{dM} \right| dM. \label{eq:ps} \ee

\noindent

\begin{figure*}
\setlength{\unitlength}{1cm} \centering
\begin{picture}(20,10)
\put(-1., -11.){\includegraphics{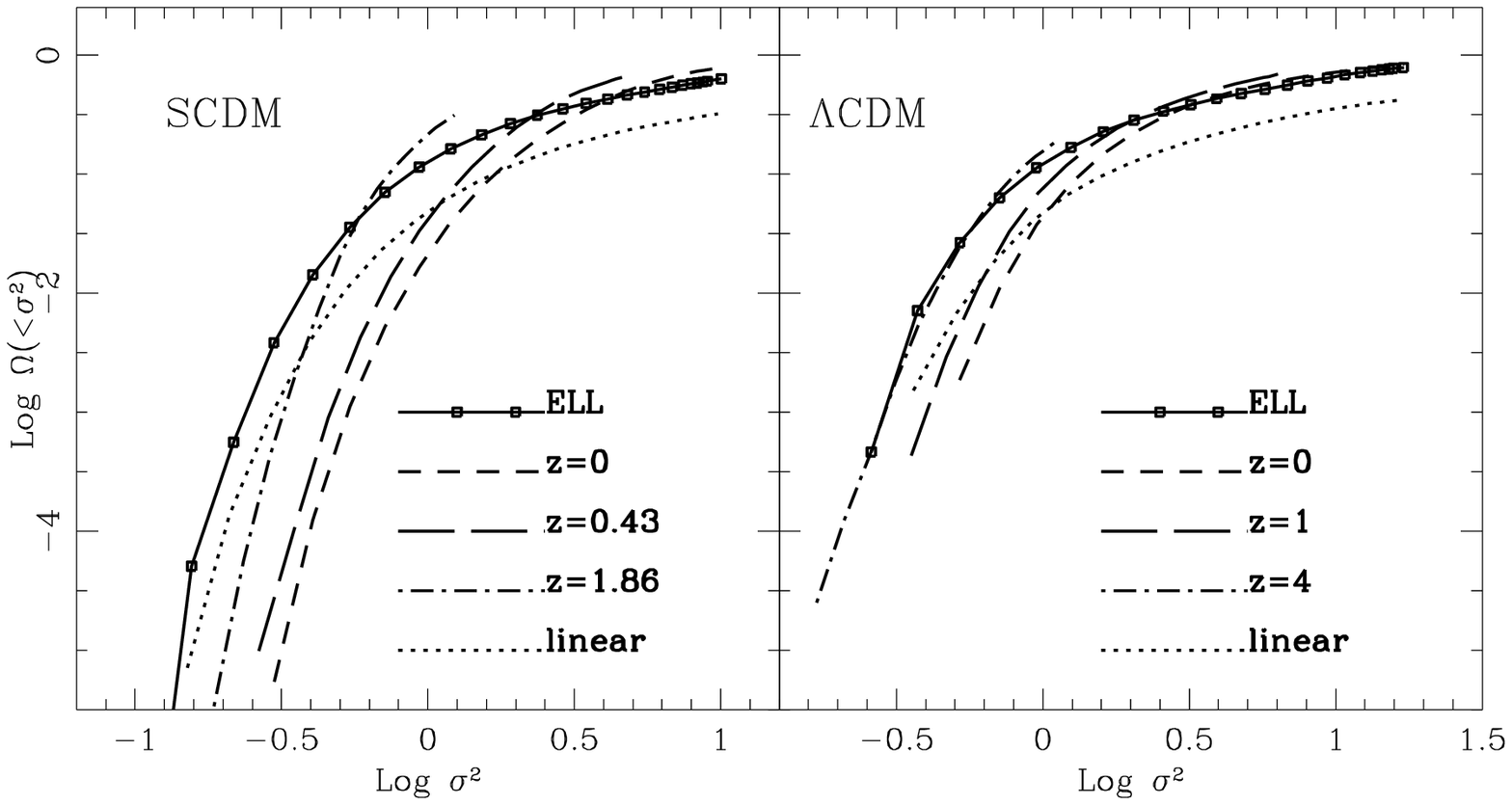}}
\end{picture}
\caption{Cumulative distributions $\Omega(<\sigma^2)$ of the $R_c$
  fields for the SCDM and \LCDMd\ simulations.}
\label{fig:omega_rc}
\end{figure*}

The functions $\Omega(<\sigma^2)$ from ELL and the simulations are
compared in figure~\ref{fig:omega_rc}. The LPT curves are by
construction independent of time and cosmology, so that only the $z=0$
LPT prediction in shown. In contrast, the $\Omega$ curves obtained
from the simulations change with time. At late times, particles have
crossed the structure they belong to many times and the numerical
displacements differ more and more from the LPT ones. This is
confirmed by the fact that the point of intersection between the
$\Omega(<\sigma^2)$ obtained from LPT and simulation roughly scales as
$b(t)^2$.  Most notably, the difference between predictions and
simulations tends to vanish for the highest redshifts; in this case
the particles have not had time to cross the structures, and their
trajectories are very similar to the LPT ones.  In all cases we notice
that the numerical $\Omega(<\sigma^2)$ functions become larger than
the LPT ones at the smallest, unsmoothed scales, especially in the
SCDM case and at higher redshift.  This is most likely due to
numerical noise present in the simulation, that enhances the level of
non-linearity of the displacements, and in the SCDM case to the
resampling from 360$^3$ to 256$^3$ grids.

For comparison, we show in figure~\ref{fig:omega_rc} also
$\Omega(<\sigma^2)$ from linear theory with $\delta_c=1.686$, which
falls short of both the ELL prediction and the simulations. We have
verified that linear theory (with Gaussian smoothing!) misses the
collapse of many mass points that belong to filaments or to low mass
halos. Decreasing $\delta_c$ to 1.5 improves the agreement only at the
largest masses, but does not solve the problem at small masses. The
Zel'dovich approximation severely under predicts $\Omega(<\sigma^2)$ at
large masses, but approaches the ELL curve for lower mass (Monaco
1997a). Consequently, using either linear theory or the Zel'dovich
approximation instead of ellipsoidal collapse, would significantly
decrease the accuracy of PINOCCHIO

We have also computed the $R_c$ field using full 3rd-order LPT. With
respect to ELL, the fraction of collapsed points increases at small
scales $R$, but decrease at large radii, because of the already
mentioned inability of 3rd-order LPT to reproduce the spherical limit
correctly (Monaco 1997a).  We have verified that the correlation with
the numerical $R_c$ field is noisier, and that the additional
small-scale contribution of collapsed matter consists mainly of
particles in filaments. Moreover, the computation is much more
demanding than the ELL case. We conclude that there is no advantage in
using the full 3rd-order LPT solution.

\subsection{$R_c$ and the simulated halos}

Having demonstrated the ability of LPT in predicting collapse in the OC
sense (without free parameters), we need to decide whether OC may be of
any use to predict which mass elements are going to end up in relaxed
halos. In order to do so, we compute the \lq mass field\rq~ from the
simulation, which assigns to each grid vertex in the initial conditions, the
mass of the halo that the corresponding particle ends-up in. Halos have
been identified in the simulation using a standard friends-of-friends
(FOF) algorithm, with a linking length 0.2\footnote{
The simulation halos were identified using a standard FOF
algorithm with linking length 0.2, irrespective of cosmology.  In this
way, halos are defined above a fixed fraction of the {\em mean}
density -- as opposed to above a fixed fraction of the {\em critical}
density. Jenkins et al. (2001) showed that this makes the mass
function almost universal with cosmology, and in addition it is
similar to the definition used in PINOCCHIO.}
times the mean inter particle distance.  The mass field is shown in
figure~\ref{fig:fields}c for the same slice of the \LCDMd\ simulation
as the other panels.  A FOF halo looks like a plateau, with the
plateau's height giving the halo's mass. There is a broad agreement
between the peaks in the $R_c$ and mass fields, because massive (low
mass) objects are generally associated with large (small) smoothing
radii. Consequently, there certainly is some connection between orbit
crossed regions and relaxed halos.  However, there are some important
differences as well.

Not all the FOF points fall within the boundaries of the $R_c$
contours. This fact was already addressed by Monaco \& Murante (1999),
and is expected because the OC criterion tends to miss those infalling
particles that have not made their first crossing of the structure. In
fact, strictly speaking those should not be counted as belonging to the
relaxed halo anyway. For SCDM, the fraction of FOF particles not
predicted to be OC-collapsed ranges from $\sim$10 per cent at large masses to
$\sim$20 per cent at smaller masses; smaller values are obtained for \LCDMd,
where the fraction of collapsed mass is higher. This has a modest
impact on the results, and is hardly noticeable in
figure~\ref{fig:fields}.

\begin{figure*}
\centerline{
\psfig{figure=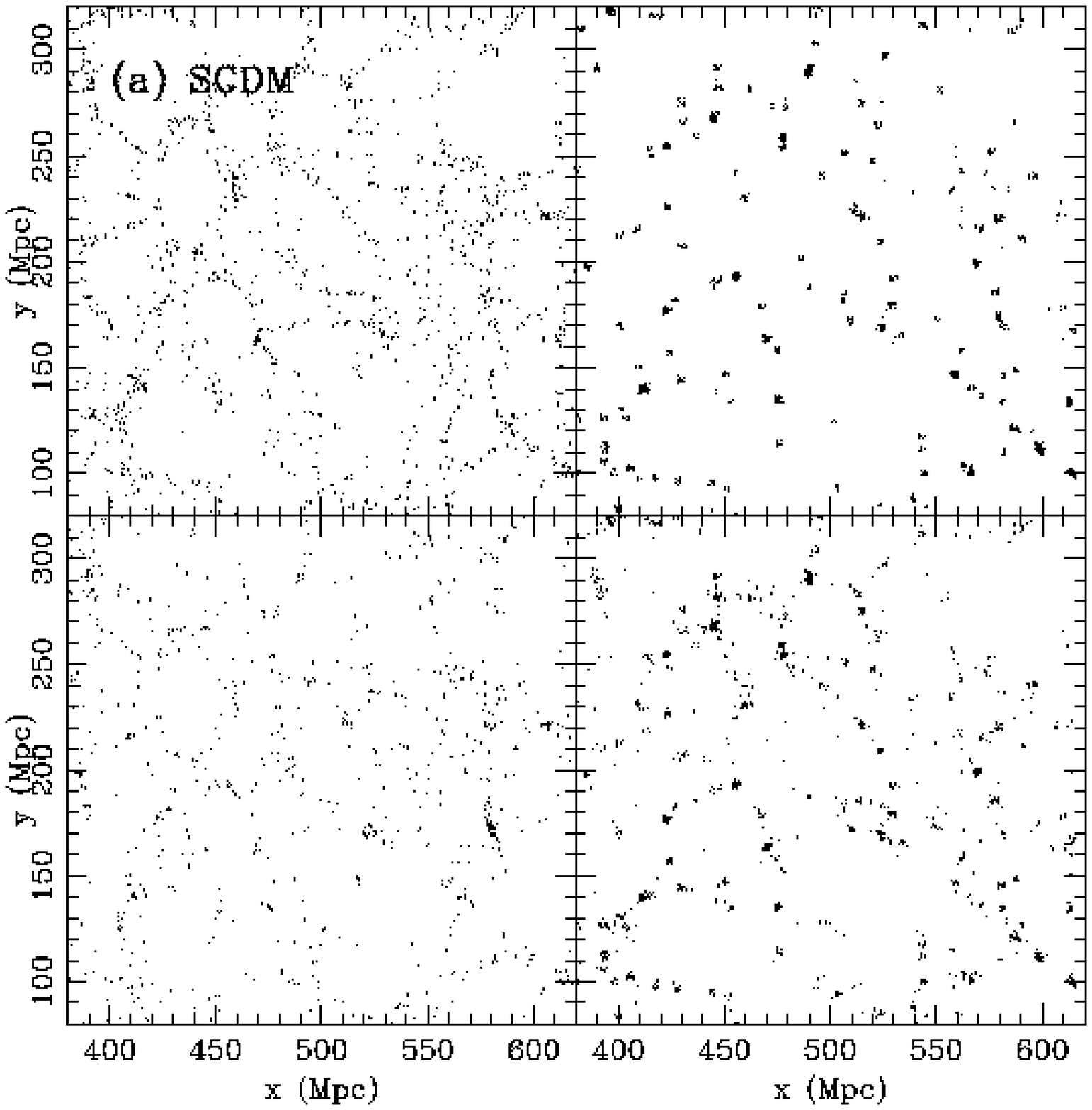,width=9.5cm}
\psfig{figure=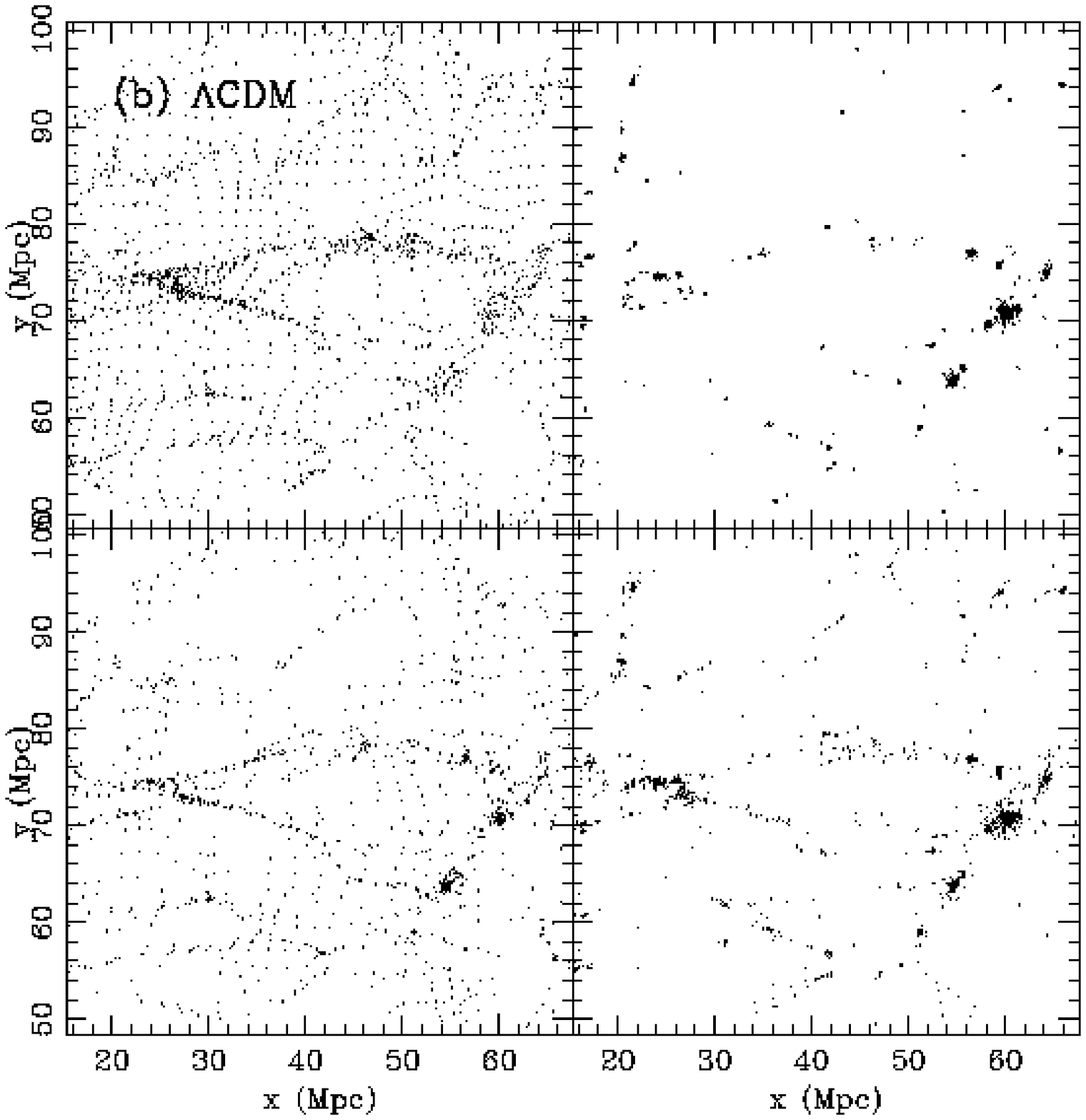,width=9.5cm}}
\caption{Final positions of particles at $z=0$ from a slice of the
initial conditions for the SCDM model (plot a) and the \LCDMd\ model
(plot b). In each plot, the left panels show those particles that are
in filaments (i.e. that have undergone OC but are not assigned to a
halo), the right panels show particles that are assigned to
halos. Upper panels are obtained from the simulations, lower panels
refer to PINOCCHIO.The large visual difference between the two
cosmologies is mostly due to the very different box size used.}
\label{fig:fil}
\end{figure*}

More importantly, the reverse is true as well: many particles assigned
non-vanishing or even high $R_c$ values do not belong to a halo.  These
particles are in the moderately over dense filaments and sheets that
connect the relaxed halos. These structures, although indeed in the
multi-stream regime, are in a relaxation state very different from that
of the halos. It is apparent that the removal of such sheets and
filaments (hereafter referred to as filaments) is an important issue
that needs to be addressed.

\subsection{Computing the collapse time}

Another feature apparent when comparing the mass and $R_c$ fields
(figure~\ref{fig:fields}) is that many FOF halos may correspond to a
single broad peak of $R_c$.  This makes the time-dependent $R_c$ field
unsuitable for addressing the fragmentation of matter into halos and
filaments.  It is more convenient to follow a procedure similar to the
merging cell model (Rodrigues \& Thomas 1996; Lanzoni et al. 2000),
i.e.  recording for each mass point the largest $F$-value it reaches,
or, in other words, the highest redshift at which the point is
predicted to collapses in the OC sense (for SCDM it is simply
$F=(1+z_c)$, where $z_c$ is the collapse redshift).  This is another
way to solve the so-called cloud-in-cloud problem (Bond et al. 1991): a
point that collapses at some redshift is assumed to be collapsed at all
lower redshifts.  We therefore record the following quantity:

\be F_{\rm max}({\bf q}) \equiv \max_R [F({\bf q}; R)]. \label{eq:fmax} \ee

\noindent
Together with \fmax\, we also store for each point the smoothing radius
\rmax\ at which $F=F_{\rm max}$, and the corresponding Zel'dovich
velocity \vmax\ computed at the time $b(t)=1/F_{\rm max}$ appropriate
for the smoothing radius \rmax.

In contrast to $R_c$, the inverse collapse time \fmax\ evidently does
not depend on time, while it does depend on the smoothing radius. The
excursion set of those points where \fmax\ is greater than some level
$F_c$ gives the mass that has collapsed before the time $t_c$ that
corresponds to $F_c$, at the highest resolution on the grid (i.e.
without smoothing, $R=0$).  The lower right panel of
figure~\ref{fig:fields} plots the \fmax\ field for the same section as
the other panels. Within each large object identified in the mass
field, $F_c$ has many small peaks that correspond to objects forming
at higher redshifts. These peaks are modulated by modes on a larger
scale that follow the excursions of the $R_c$ field. Those large scale
modulations are ultimately responsible for the later merging of these
small peaks into the massive object identified at late times. In this
way, PINOCCHIO combines the information on the progenitors to
reconstruct the merger history of objects, as described in detail in
the next section.

\section{Identification and merger history of halos}

In the PS and excursion set approaches the mass of the objects that
form at a scale R is simply estimated as 

\be M \simeq \frac{4\pi}{3} \bar{\rho} R^3. \label{eq:mass} \ee

\noindent
A more detailed treatment of the complex processes that determine the
shape of the Lagrangian region to collapse into a single halo is
required to get an improved description of the formation of the
objects, and thus an improved agreement with simulations at the
object-by-object level. In PINOCCHIO, this is done by generating
realisations of the density field on a regular grid, computing the
\fmax\ field as explained above, and then \lq fragmenting\rq~ the
collapsed medium into halos and filaments by considering the fate of
each particle separately. To enable a detailed comparison with the
simulation, we will perform these steps on the initial conditions of
the runs. Therefore, we can compare the properties of {\em individual
halos} between the simulations and PINOCCHIO, not just the statistics
of halos. Of course. PINOCCHIO can be applied to any realisation of a
density field, including non-cubic grids and non-Gaussian perturbation
fields. The fragmentation algorithm can even be applied to non-regular
and non-periodic grids, if the FFT-based calculation of \fmax\ is
suitably modified.

\begin{figure*}
\centerline{
\psfig{figure=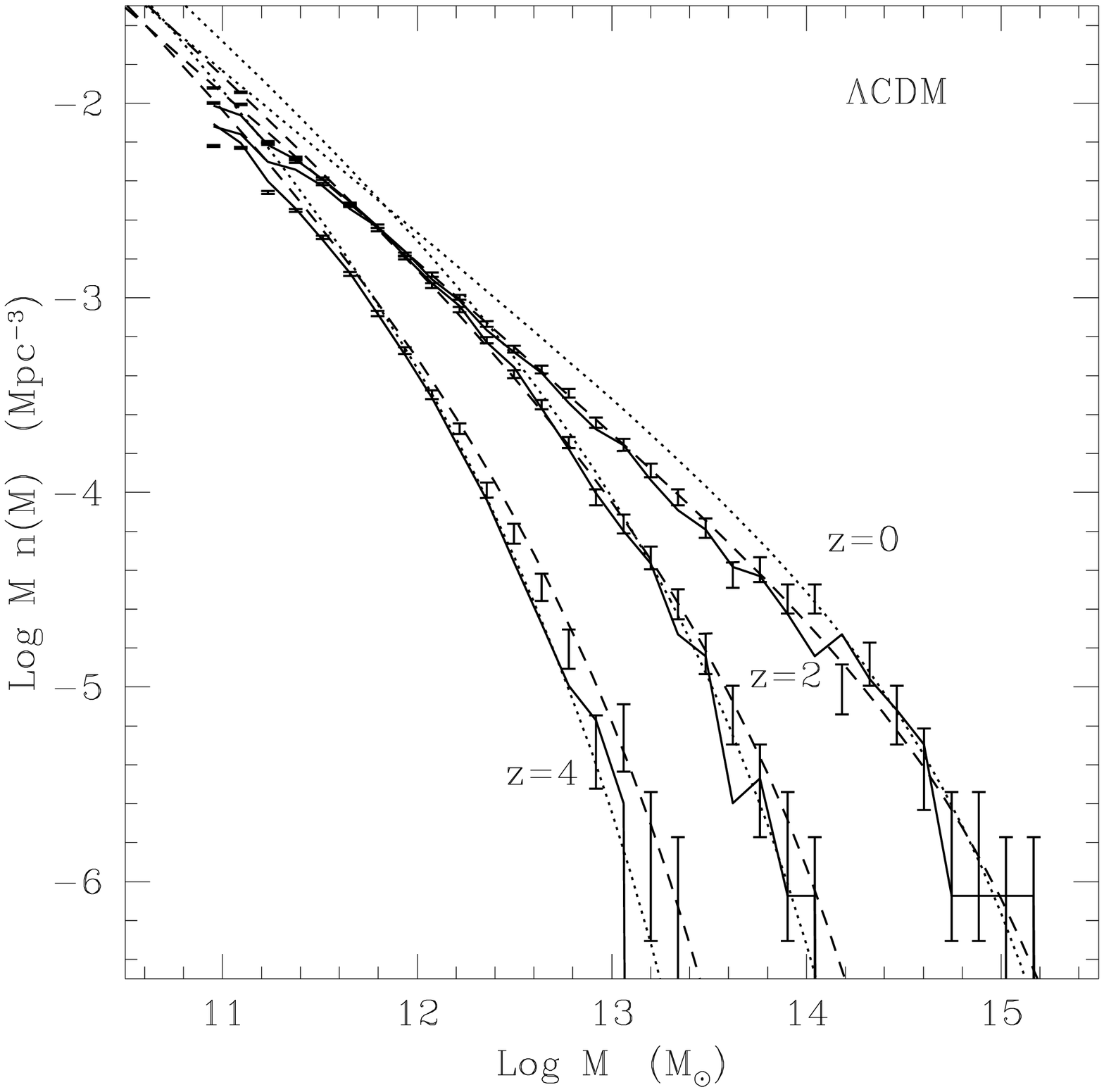,width=18cm}}
\caption{
  Mass functions for the \LCDMd\ model at different redshifts
  indicated in the panel.  Error bars denote Poissonian errors for
  the simulated mass function, continuous lines are the PINOCCHIO
  predictions, dotted and dashed lines are the PS and ST predictions,
  respectively.}
\label{fig:mf}
\end{figure*}

The fragmentation code mimics the two main processes of hierarchical
clustering, that is the accretion of mass onto halos and the merging
of halos.  The particles of the realisation are considered in order of
descending \fmax-value, i.e.  in chronological order of collapse.  At
a given time the particles that have already collapsed will be either
assigned to a specific halo, or associated with filaments. Because of
the continuity of the transformation between Lagrangian and Eulerian
coordinates, equation~\ref{eq:lag}, a particle must touch a halo in
the Lagrangian space if it is to accrete on it.\footnote{Here it is
  assumed that a particle that accretes onto a halo never escapes back
  in to the field. Such stripping does occasionally happen in
  simulations, but not very often and we neglect it.}  Thus a
collapsing particle can accrete only onto those halos that are \lq
touched\rq\ by it, i.e. that already contain one of its 6 nearest
neighbours in the Lagrangian space of initial conditions (we call
these particles {\em Lagrangian neighbours}).  To decide whether the
particle does accrete onto a touching halo, we displace it to the
Eulerian space according to its \vmax\ velocity.  The halo is
displaced to its Eulerian position at the time of accretion, using the
average velocity of all its constituent particles\footnote{Thus the
  velocity of a halo is an average over velocities calculated at
  different smoothing radii.  A better estimate (but expensive in
  terms of computer memory) would be to average the unsmoothed
  velocities over the particles of the halo. Fortunately,the stability
  of the velocity to smoothing makes the two estimates very similar,
  once the average is performed over many particles.}.  
In the following we express sizes and distances in terms of the grid
spacing.  The size \rlag\ of a halo of $N$ particles is taken to be

\be R_{\rm N} = N^{1/3}. \label{eq:rlag} \ee

\noindent
The collapsing particle is assumed to accrete onto the halo, if the
Eulerian (comoving) distance $d$ between particle and halo is smaller
than a fraction of the halo's size \rlag\,

\be d < f_{\rm a} \times R_{\rm N}. \label{eq:accr} \ee

\noindent
The free parameter $f_{\rm a}$, which is smaller than one, controls
the over density that the halo reaches in the Eulerian space,
$1+\delta_{\rm halo} \sim 3/4\pi f_{\rm a}^3$.  Therefore, this
criterion selects halos at a given over density, making it similar to
the usual FOF or similar selection criteria.  The value of the $f_{\rm
a}$ parameter is fixed in Appendix A to $\sim 0.25$.  Then, the halos
reach a much lower over density than the value $\sim 200$ used in
simulations; Zel'dovich velocities (and LPT velocities in general) are
not accurate enough to reproduce such high densities for the relaxed
halos. However, PINOCCHIO only attempts to identify the halos, not
compute their internal density profile as well.

\begin{figure*}
\setlength{\unitlength}{1cm} \centering
\begin{picture}(20,13)
\put(-1., -9.){\includegraphics{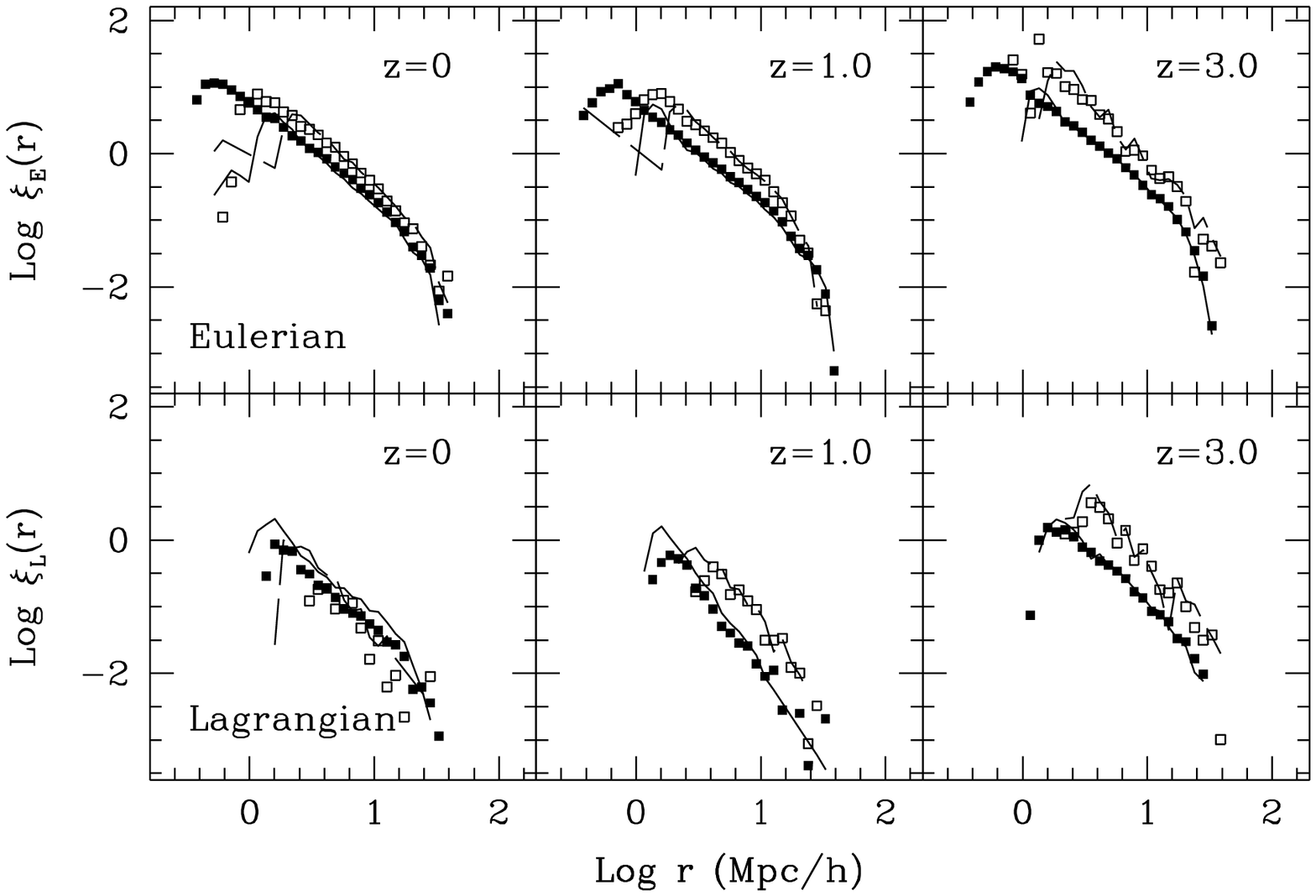}}
\end{picture}
\caption{Eulerian correlation function (upper panels) and Lagrangian
correlation function (lower panels) for the \LCDMd\ models at three
redshifts indicated in the panel, and for two mass ranges.  Symbols
refer to simulation results, lines to PINOCCHIO predictions.  Filled
squares and continuous lines: correlation function for low mass halos
(mass $M$ from $6.3 \times 10^{11}$ to $3 \times 10^{12}\ M_\odot$),
open squares and dashed lines: correlation function for massive halos
($M > 3 \times 10^{12}\ M_\odot$).}
\label{fig:xi}
\end{figure*}

When a collapsing particle touches two (or more) halos in the
Lagrangian space, then we use the following criterion to decide whether
the two halos should merge. We compute the Eulerian distance $d$
between the two halos at the suspected merger time using the halo
velocities described above. The halos are deemed to merge when $d$ is
smaller than a fraction of the Lagrangian radius of the larger halo:

\be d < f_{\rm m} \times \max (R_{\rm N1},R_{\rm N2}). \label{eq:merg} \ee

\noindent
This condition amounts to requiring that the centre of mass of the
smaller halo, say halo 2, is within a distance $f_{\rm m} R_{\rm N1}$
of the centre of mass of the larger halo 1.  The value of the $f_{\rm
m}$ parameter is fixed in Appendix A to 0.35.

We note that PINOCCHIO is not restricted to binary mergers. In
principle, a particle has 6 Lagrangian neighbours so up to 6 halos may
merge at the same time. In practice binary mergers are the most
frequent, but ternary mergers also occur, while mergers of four halos
or more are rare.

In more detail, the fragmentation code works as follows.  We keep track
of halo (or filament) assignment for all particles.  For each
collapsing particle we consider the halo assignment of all Lagrangian
neighbours; touching halos are those to which a Lagrangian neighbour
has been assigned.  The following cases are considered:

(i) If none of the neighbours have collapsed, then the particle is a
local maximum of \fmax.  This particle is a seed for a new halo of unit
mass, created at the particle's position.

(ii) If the particle touches only one halo, then the accretion
condition is checked.  If it is satisfied, then the particle is added
to the halo, otherwise it is marked as belonging to a filament.  The
particles that only touch filaments are marked as filaments as well.

(iii) If the particle touches more than one halo, then the merging
condition is checked for all the touching halo pairs, and the pairs
that satisfy the conditions are merged together.  The accretion
condition for the particle is checked for all the touching halos both
before and after merging (when necessary). If the particle can accrete
to both halos, but the halos do not merge, then we assign it to that
halo for which $d/R_{\rm N}$ is the smaller. Occasionally, particles
fail to accrete even though the halos merge.

(iv) When a particle is accreted onto a halo, all filament particles
that neighbour it are accreted as well.  This is done in order to
mimic the accretion of filaments onto the halos.  Notice that up to 5
filament particles can flow into a halo at each accretion event.

This fragmentation code runs extremely quickly, in a time almost
linearly proportional to the number of particles. At late times,
slightly more time is spent in updating the halo assignment lists in
case of mergers, but this does not slow down the code much.

In high density regions where most of the matter has collapsed, it can
happen that pairs of halos that are able to merge are not touched by
newly collapsing particles for a long time.  This problem can be solved
by keeping track of all the pairs of touching halos that have not
merged yet, and checking the merging condition explicitly at some time
intervals.  Such a check slows the code down significantly, and has
only a moderate impact on the results when the fraction of collapsed
mass at the grid scale is large. Similarly, the accretion of filament
particles on to halos can be checked at some given time intervals, but
again, the impact is modest on the results but the increase in computer
time may be substantial

\begin{figure}
\centerline{
\psfig{figure=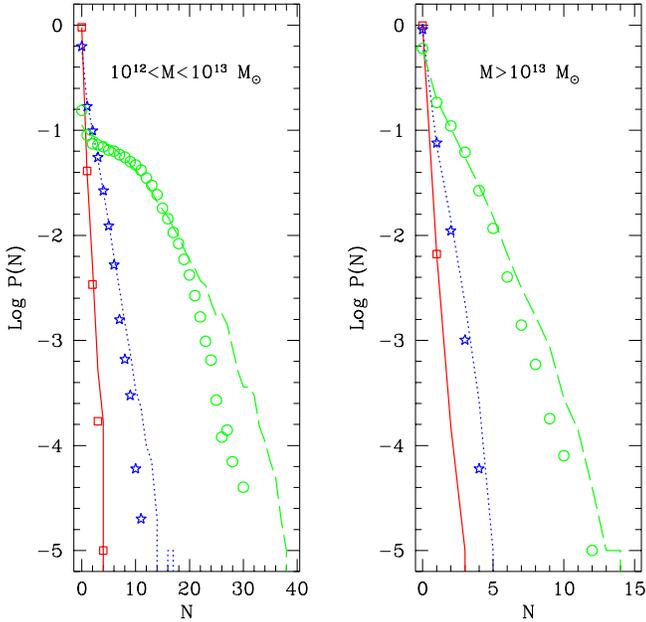,width=9cm}}
\caption{ Count-in-cell analysis of the halo catalogues at $z=0$.
Left and right panels show results for the mass ranges indicated.
Symbols refer to simulation results, lines to PINOCCHIO predictions.
Continuous, dotted and dashed lines (or squares, stars and circles)
refer to cell sizes of 2, 5 and 10 Mpc (1.3, 3.25 and 6.5 Mpc/h).}
\label{fig:counts}
\end{figure}

While the dynamical estimate of collapse time does not introduce any
free parameter, the fragmentation process does.  The same happens in
the simulation, where any halo-finding algorithm has at least one free
parameter, such as the linking length for FOF halos. This is because
the definition of what constitutes a DM halo is somewhat arbitrary,
and hence also the corresponding mass function is not unique (Monaco
1999).  Fortunately, different clump-finding algorithms usually give
similar results, so that this ambiguity is in general not a real
problem. In the following, the best fit parameters for PINOCCHIO will
be chosen so as to reproduce the mass function of the FOF halos of the
simulations, with linking length equal to 0.2 times the inter-particle
distance, at many redshifts.  We have checked with one SCDM output
that the differences in the halos as defined by the HOP (Eisenstein \&
Hut 1998) and SO (Lacey \& Cole 1994) algorithms are much smaller than
the accuracy with which we are able to recover the FOF halos.

The five free parameters of the fragmentation code, and the
determination of their best-fit values, are described in Appendix A.
Note that the results shown paper I use a more limited set of three
free parameters, which is adequate to describe large-volume
realisations such as that of the SCDM simulation, while the larger set
of parameters described in Appendix A has a more general validity.

The ability of PINOCCHIO to distinguish OC particles that collapse into
halos versus those that remain in filaments, is shown in
figure~\ref{fig:fil}. In this figure we plot the final position of the
particles, as given by the simulation output at redshift $z=0$, for a
section of the initial conditions of the SCDM and \LCDMd\ simulations.
Left panels show only the filament particles, defined as those which
are in OC according to $R_c$ but do not belong to any halo.  Right
panels show only those particles that are in halos.  Upper panels show
the result from the simulation, lower panels the PINOCCHIO
predictions. Clearly, PINOCCHIO is able to distinguish accurately halos
from filaments, even though some filament particles are interpreted as
halo particles and vice versa.  When compared with figures 6 and 7 of
Bond et al. (1991), figure~\ref{fig:fil} shows the marked improvement
of PINOCCHIO with respect to the extended PS approach. We want to
stress that filaments are important in their own right. For example,
most of the Lyman-$\alpha$ absorption lines seen in the spectra of
distant quasars are produced in filaments (e.g. Theuns et al. 1998), so it
will be useful to be able to generate catalogues of halos {\it and}
filaments.

\section{Detailed comparison to simulations.}

\subsection{Statistical comparison}

The comparisons of PINOCCHIO and FOF mass and correlation functions
for the SCDM simulation were presented in paper I, using the more
limited set of three free parameters.  The results with the full
five-parameter set are very similar and are not shown here.  In
figure~\ref{fig:mf}, we compare the mass function computed using
PINOCCHIO and the \LCDMd\ $N$-body simulation.  The FOF halos were
identified as explained above.  For reference, we also plotted the PS
and Sheth \& Tormen (1999, hereafter ST) mass functions. The choice of
parameters reported in Appendix A produces a PINOCCHIO mass function
which falls to within $\sim$5 per cent of the simulated one from $z=5$
to $z=0$, for all mass bins with more than 30-50 particles per halo
and for which the Poisson error bars are small.  The only residual
systematic is a modest, $\sim$10-20 per cent underestimate at the
highest-mass bins and highest redshift.  An accuracy of better than 10
per cent on the mass function for a given realisation is perfectly
adequate for most applications, as it is usually smaller than the
typical sample variance as well as the intrinsic accuracy of
$\sim20-30$ per cent with which the mass function of N-body
simulations is defined.  Because PINOCCHIO is calculated for the {\em
same} initial conditions as the simulation, Poisson error bars are not
the correct errors to use for this comparison (notice that the Poisson
error bars of the PINOCCHIO mass function are obviously very similar
to those of the numerical one).  We show them both for comparison with
PS and ST and to understand which mass bins are affected by small
number statistics.

\begin{figure*}
\setlength{\unitlength}{1cm} \centering
\centerline{
\psfig{figure=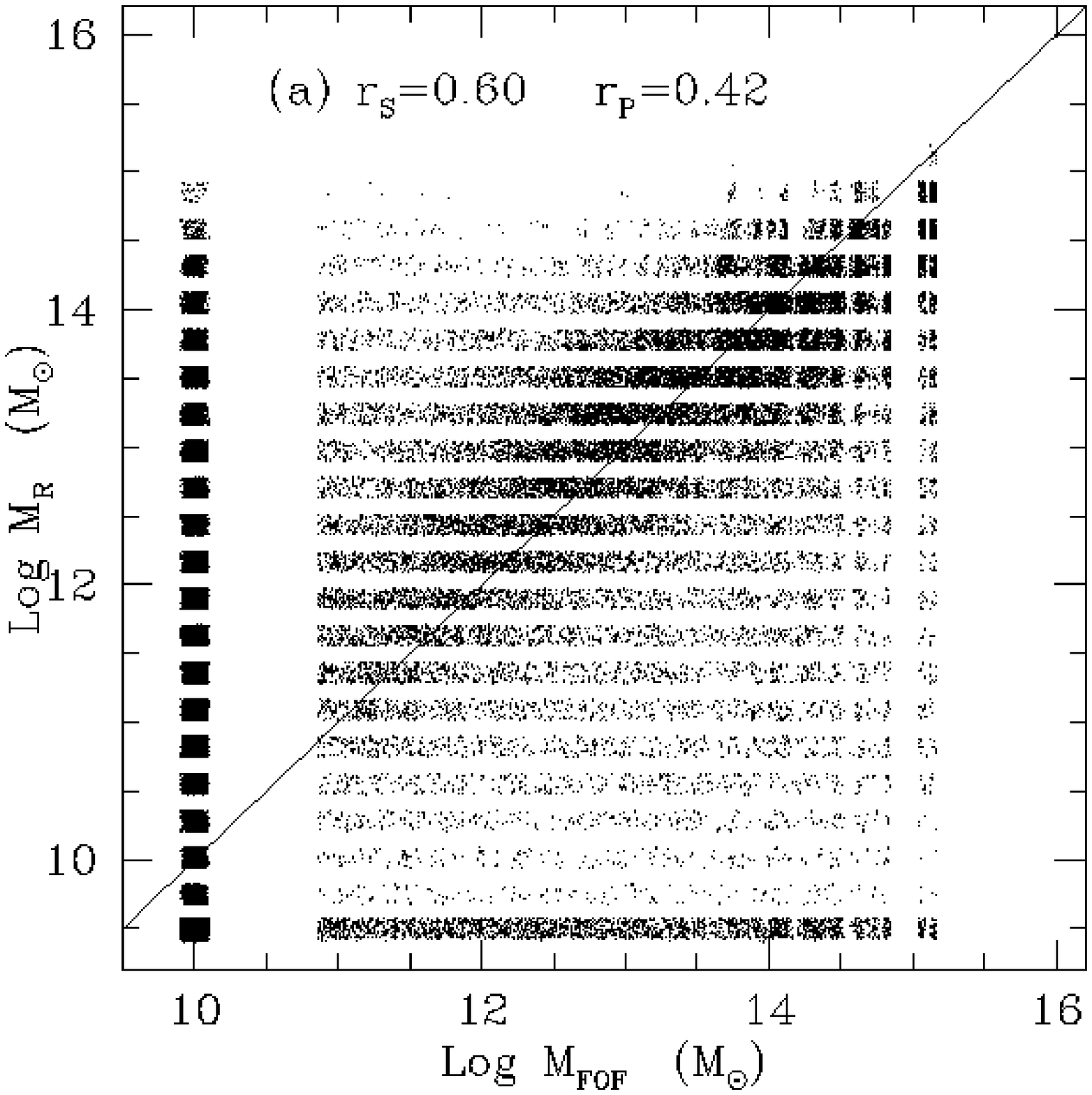,width=9.cm}
\psfig{figure=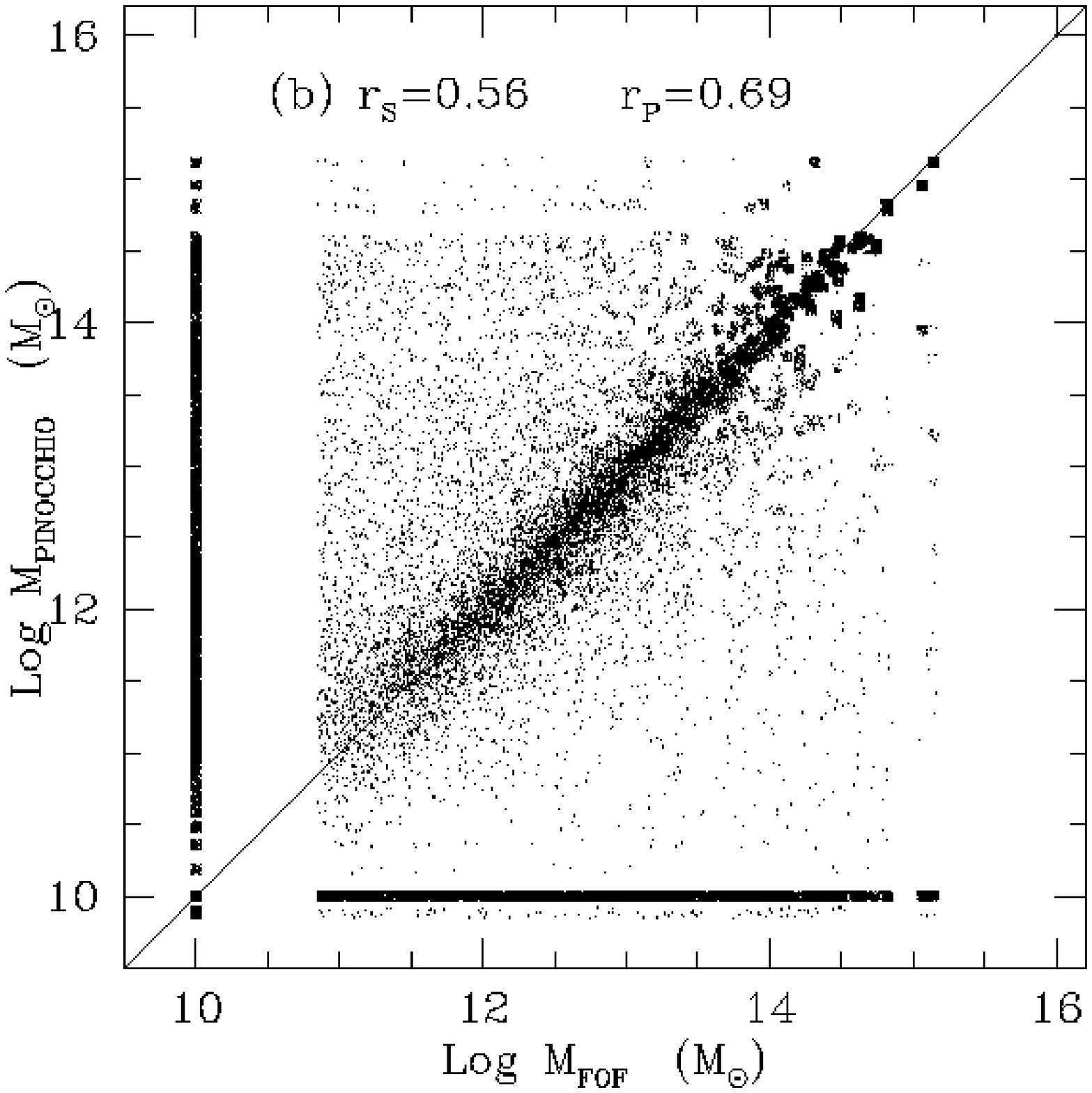,width=9.cm}}
\caption{Comparison of the mass fields for FOF halos identified in the
\LCDMd\ simulation with that obtained from the $R_c$ field using the
PS mass-radius relation (equation~\ref{eq:mass}, left panel) and using
PINOCCHIO (right panel).  For clarity some random noise has been added
to the mass field values, especially to that obtained from the
discrete $R_c$ field.}
\label{fig:mass}
\end{figure*}

Taking the ST mass function (or the analytic fit of Jenkins et al.
2001) as a {\em bona fide} estimate, we have checked the validity of
PINOCCHIO in reproducing the mass function of halos in a wide variety
of cosmologies and box sizes (see Appendix A).  The fit of the mass
function is found to be still good even for halo masses as small as
$10^5$ M$_\odot$ ($\Lambda$CDM cosmology), at a redshift high enough
to avoid that the whole box goes non-linear.

Strictly speaking the agreement between the mass functions of
PINOCCHIO and the simulated ones is not a proper comparison of
prediction with numerical experiment, as the fit is achieved by tuning
the free parameters discussed in Section 3.  However, the very
existence of a limited set of parameters that allows to achieve such a
good agreement in different cases (SCDM and $\Lambda$CDM, PKDGRAV and
Hydra, small and large boxes) is a very important result.  As shown
also in figure~\ref{fig:mf}, PINOCCHIO improves the fit with respect
to PS, giving an accuracy very similar to the ST fit.  Jenkins et
al. (2001) showed that the ST fit underestimates the knee of the FOF
mass function by $\sim$10-20 per cent\footnote{ Sheth \& Tormen (2001)
show that a modest tuning of their parameters can remove this
disagreement.}; we have verified that when this difference is evident
the best fit PINOCCHIO mass function is more similar to the numerical
one and to the Jenkins et al. (2001) fit than to the ST mass function.
This is evident in figure 1 of paper I (where the residuals of the
$z=0$ mass functions are shown), but is hardly noticeable in
figure~\ref{fig:mf}, where Poisson errorbars are larger.  The
comparison with the \LCDMd\ simulation shows that the fit is very good
also down to the low mass tail $M\sim 10^{11}\ M_\odot$ or $M/M_*\sim
10^{-2}$ ($M_*$ denotes the characteristic mass of the PS mass
function, such that $\sigma^2(M_*)=\delta_c^2$; see, e.g., Monaco
1998).

In the PS and excursion set approaches, the mass function is \lq
universal\rq\ when expressed in terms of the variable
$\Omega(<\sigma^2)$ already defined in Section 2.2
(equation~\ref{eq:ps}), which in this case gives the fraction of mass
collapsed into objects larger than $M(\sigma^2)$ (with the mass given
by equation~\ref{eq:mass}). The mass functions obtained from a large
set of numerical simulations is indeed found to be universal to within
$\sim$30 per cent (Jenkins et al.  2001).
The PINOCCHIO mass function is not by construction universal, yet we
find it to be nearly universal once the resolution effects described in
Appendix A are taken into account.

However, the mass function of the Governato et al. (1999) SCDM
simulation used here shows an excess of massive halos at high
redshift. This was already noticed by Governato et al., and quantified
as a drift of the $\delta_c$ parameter from $\sim$1.5 at high redshift
to $\sim$1.6 at $z=0$.  This trend is not confirmed by other
simulations (Jenkins et al. 2001), nor by our \LCDMd\ simulation
presented here.  We find that PINOCCHIO reproduces the weak trend of
Governato et al. (1999) in the SCDM simulation, but also the lack of
such a trend in the \LCDMd\ one.  We conclude therefore that this
effect is likely to be linked to the initial conditions generator,
which is different for the two realisations (see Appendix A for more
details). Recall that the PINOCCHIO mass functions refer to the {\em
same} initial conditions as were use to perform the simulations.

In figure~\ref{fig:xi} we show the correlation function of halos as a
function of mass, both in Eulerian and in Lagrangian space. The
correlation function has been computed using a standard pair counting
algorithm. The agreement between PINOCCHIO and the simulation is very
good down to scales of a few grid cells, i.e. $\sim$1-2 comoving Mpc/h
(larger for rarer objects), below which the PINOCCHIO correlation
functions become negative.  This is in agreement with what found in
paper I for the SCDM simulation.  The differences are of order
$\sim$10-20 per cent in amplitude and $\la$10 per cent in terms of
scale at which a fixed amplitude is reached. This means that both the
correlation length $r_0$, at which $\xi(r_0)=1$, and the length at
which $\xi=0$ are reproduced with an accuracy of better than 10 per
cent.  This is an improvement with respect to the ST formalism, where
the accuracy is of order $\sim$20 per cent (Colberg et al. 2001).
More importantly, the trends of increased correlation for the more
massive halos, or for halos of a given mass with increasing redshift,
are both well reproduced. The correlation functions in the Lagrangian
space are noisier, and are reproduced with somewhat larger error,
especially at $z=0$ where they are slightly overestimated; however
this error does not seem to propagate to the Eulerian correlation
functions.

The two-point correlation function gives only a low-order statistics
of the spatial distribution of a set of objects.  To probe the
accuracy of the PINOCCHIO results at higher orders, we have performed
a count-in-cell analysis of the halo distribution, which, at variance
with the correlation function, depends also on the phases of the space
distribution of the halos.  This is shown in figure~\ref{fig:counts}
for galactic-sized ($10^{12} M_\odot \le M \le10^{13} M_\odot$) and
group-sized ($M \ge 10^{13} M_\odot$) halos of the \LCDMd\
realisation, and cell sizes of 2, 5 and 10 Mpc (corresponding to 1.3,
3.25 and 6.5 Mpc/h).  The count-in-cells curves are well reproduced by
PINOCCHIO, although their skewness is slightly underestimated,
especially for larger cells and smaller masses.  In particular, the
void probability $P_0$ of finding no halos in the cell is reproduced
with an accuracy no worse that a few percent when it takes values in
excess of 0.6.

\subsection{Point-by-point and object-by-object comparison}

The PINOCCHIO approach is not just limited to making accurate
predictions for statistical quantities such as the mass and
correlation functions, but is also able to predict halo properties
that correspond in detail to those obtained from simulations. This is
in contrast to the PS approach, where the object-by-object agreement
is very poor (White 1996, but see Sheth et al. 2001 for a different
view).

\begin{figure}
\setlength{\unitlength}{1cm}
\centering
\begin{picture}(20,18)
\put(-1, -3.5){\includegraphics{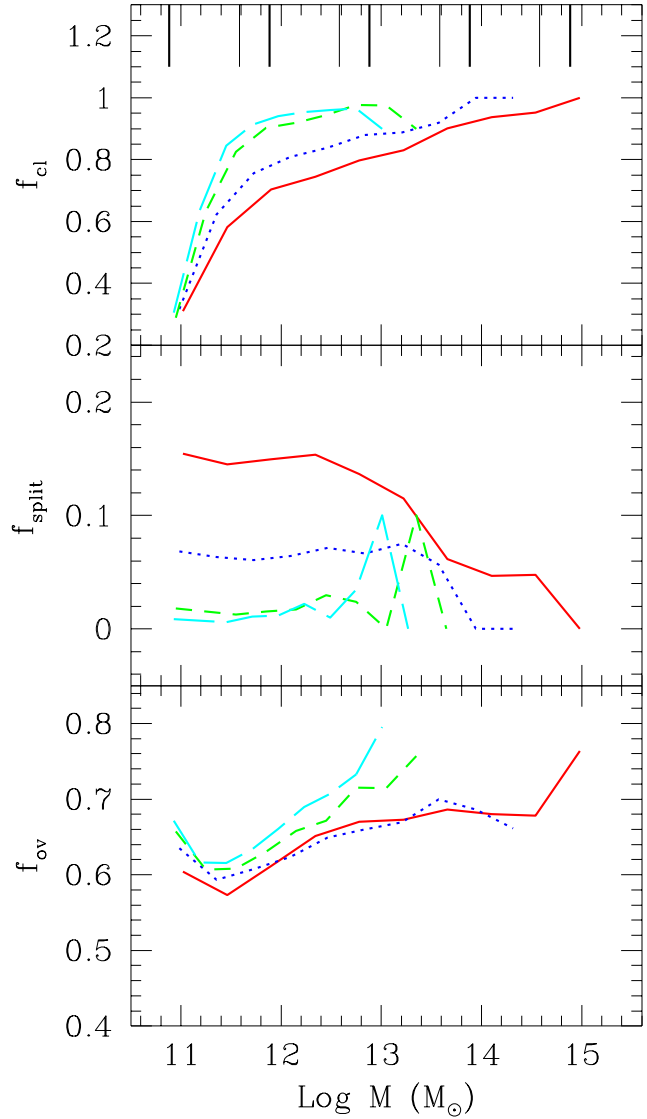}}
\end{picture}
\caption{Comparison on an object-by-object level of halos identified
by PINOCCHIO and found in the \LCDMd\ simulation, using a variety of
statistics.  Continuous, dotted, short-dashed and long-dashed lines
refer respectively to redshifts $z=0$, 1, 2 and 4.  Top panel:
fraction \fcl\ of cleanly assigned objects; middle panel: fraction
\funcl\ of non-cleanly assigned objects; bottom panel: average overlap
$f_{\rm ov}$ for cleanly assigned objects.  The vertical lines in the
top panel indicate halos with 10, 100, 10$^3$, 10$^4$ and 10$^5$
particles (heavy lines) or 50, 500, 5$\times$10$^3$ and
5$\times$10$^4$ particles (light lines).}
\label{fig:obj}
\end{figure}

Agreement at the \lq point-by-point level\rq~ requires that each
particle is predicted to reside in the correct halo with the correct
mass.  Whether this agreement holds can be checked by comparing the
mass fields already defined in section 2.2 (an example of which is
shown in figure~\ref{fig:fields}c).  We note that this type of analysis
is similar to that of Section 2.2, where the point-by-point agreement
was checked for the $R_c$ fields. In the PS approach, the mass of the
halo to which a particle belongs is estimated as in
equation~\ref{eq:mass}, with the $4\pi/3$ valid for top-hat smoothing
(or sometimes left as a free parameter).  In this case the mass field
is simply related to the $R_c$ field.  A comparison between the mass
fields obtained from the same $R_c$ field of figure~\ref{fig:rc} (with
arbitrary normalisation) and that of the simulation, $M_{\rm FOF}$,
reveals only a poor correlation, as shown shown in
figure~\ref{fig:mass} (left panel) for a random sample of $\sim$20000
particles extracted from the \LCDMd\ simulation.  The tightness of the
correlation is again quantified by the $r_S$ and $r_P$ coefficients.
This figure is similar to figure 8 of White (1996) and figure 2 of
Sheth et al. (2001), with the difference that here the $R_c$ curve
were computed with ELL instead of with linear theory (and Gaussian
smoothing instead of top-hat). The point-by-point agreement is much
better with PINOCCHIO (right hand panel), where the linear correlation
coefficients $r_P$ jumps from 0.42 to 0.69, demonstrating the increase
in accuracy.  Clearly, the improvement of PINOCCHIO in the
point-by-point comparison is not primarily due to the more accurate
dynamical description of collapse.  Rather it is due to the much more
accurate description of the {\em shape} of the collapsing region,
which is not restricted to the simple PS relation of
equation~\ref{eq:mass}.  

\begin{figure*}
\centerline{
\psfig{figure=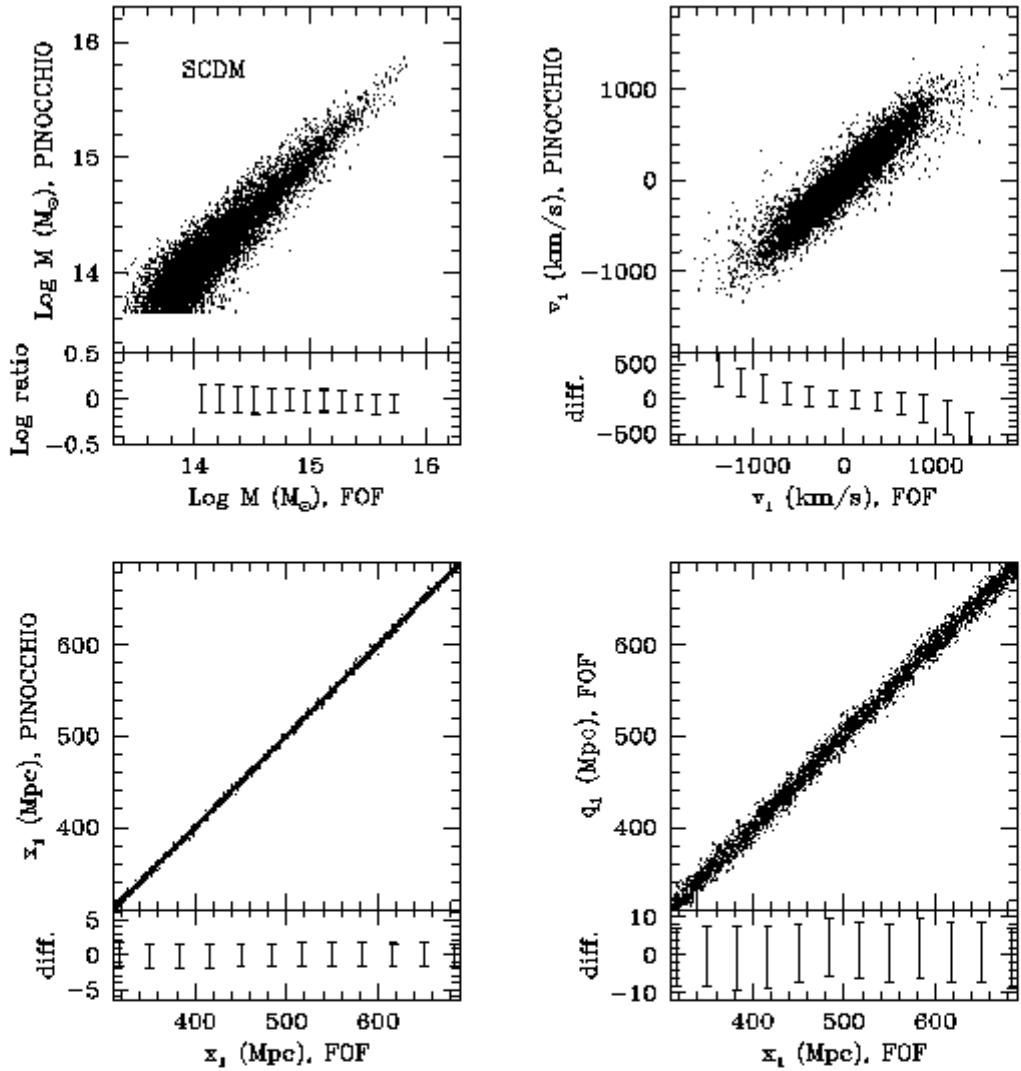,width=17cm}}
\caption{Difference in mass, position and velocity, $\log M$, $x_1$
and $v_1$ respectively, as estimated by PINOCCHIO and found from the
simulation, for cleanly assigned halos. The scatter around the mean is
plotted below each panel.  The lower right panels show for comparison
the displacement of halos according to the simulation.  The first set
of panels refer to the SCDM simulation, the second to the \LCDMd\ one}
\label{fig:scatter}
\end{figure*}

\addtocounter{figure}{-1}
\begin{figure*}
\centerline{
\psfig{figure=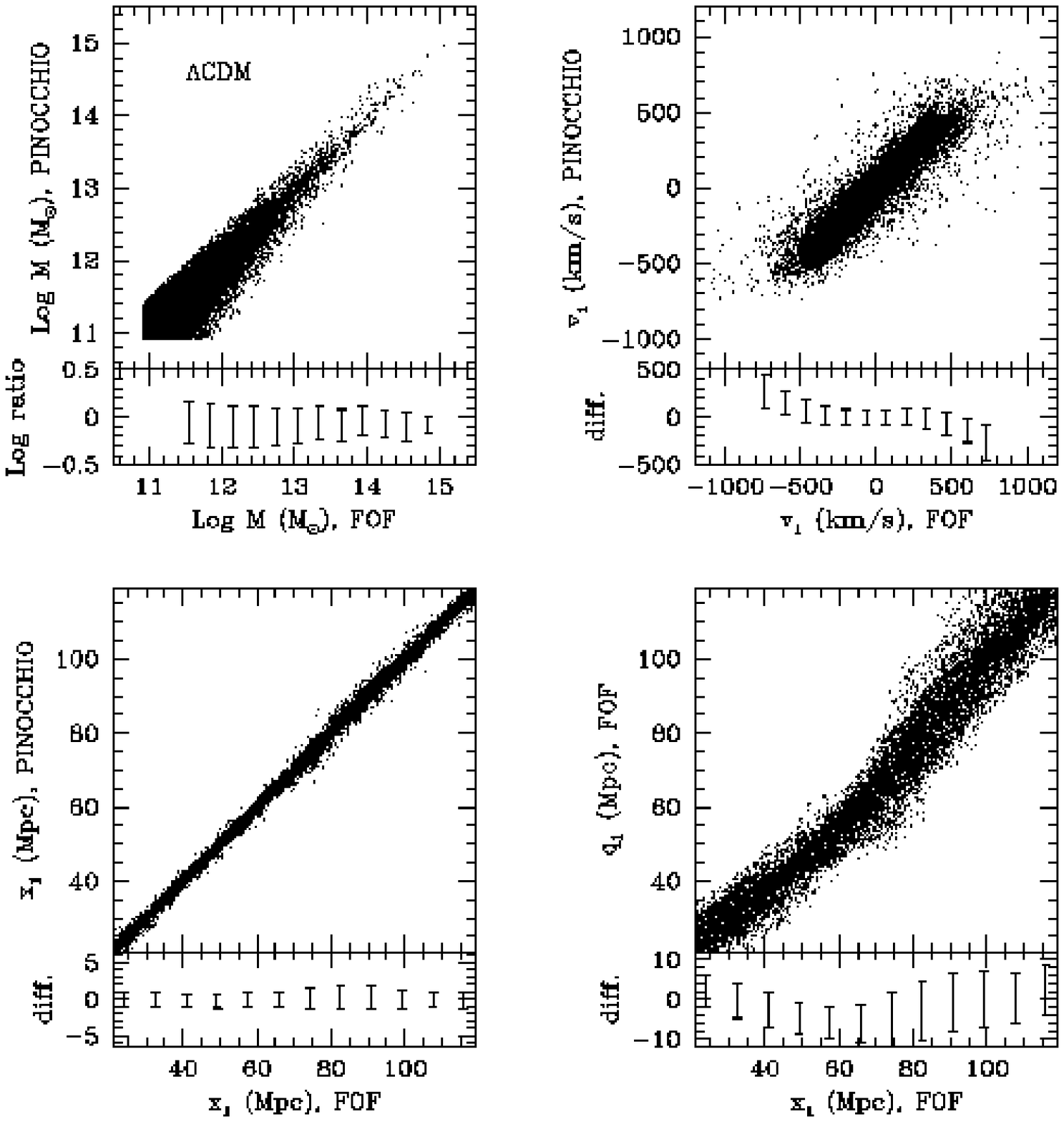,width=17cm}}
\caption{(continued)}
\end{figure*}

While the linear correlation coefficients improves significantly going
from $R_c$ to the PINOCCHIO mass field, the Spearman correlation
coefficient $r_S$ does not change much, since both panels contain a
large number of outliers.  These are particles that lie at the border
of halos, and are assigned to a halo by the simulation but not by
PINOCCHIO, or vice versa.  Such outliers are expected whenever the
boundaries of halos in the Lagrangian space are not perfectly
recovered.  On the other hand the presence of such outliers is not
very important when the catalogue of objects is considered.

We next investigate the agreement of PINOCCHIO with the simulations at
the object-by-object level, a coarser level of agreement but more
relevant in practice.  The degree of matching between halo catalogues
is quantified as in paper I.  For each object of one catalogue, the
objects of the other catalogue that overlap for at least 30 per cent
of the Lagrangian volume are considered.  Two halos from different
catalogues are \lq cleanly assigned\rq~ to each other, when each
overlaps the other more than any other halo. The fraction of halos not
cleanly assigned is \funcl.  The remainder 1--\fcl--\funcl\ is the
fraction of objects of one catalogue that does not overlap with any
halo in the other catalogue. These fractions quantify the level to
which two catalogues describe the same set of halos.  Another useful
quantity is \fov, the average fraction that halos overlap when they
are cleanly assigned.  All these estimators depend on whether
PINOCCHIO is compared with simulations or vice-versa, but in general
that difference is small as long as the comparison is good.

In figure~\ref{fig:obj} we show the values of these three indicators
of the agreement between the two halo catalogues as a function of halo
mass, for \LCDMd\ model; the SCDM case was shown in paper I.  The
agreement is very good at higher redshift with $\sim$80-90 per cent of
objects cleanly assigned when the halos have at least 50 particles.
The degree of splitting is only $\la$5 per cent, while the average
overlap of cleanly-assigned objects \fov\ ranges from 60 per cent to
70 per cent nearly independent of mass and encouragingly larger than
the 30 per cent lower limit. These results are in agreement with the
SCDM ones presented in paper I.  The agreement is slightly worse at
lower redshift, with \fcl$\ga$70 per cent for halos with at least 100
particles, and a \funcl$\sim$5--10 per cent.  Within perturbative
approaches there is obviously no advantage in going to higher
resolution, as the accuracy of LPT worsens with the degree of
non-linearity (see figure~\ref{fig:error}) and with it all the
results.  Anyway, the agreement is still very significant for the last
output, with a high fraction of cleanly assigned objects and a modest
degree of splitting.  In any case the results always improve with
increasing number of particles. Monaco (1997a) estimated that LPT
would break down when $\sim$50 per cent of the mass has undergone OC.
Therefore, the agreement shown in figure~\ref{fig:obj} (and also in
figure~\ref{fig:fil}) is better than expected.

\begin{figure*}
\setlength{\unitlength}{1cm}
\centering
\begin{picture}(20,10)
\put(-1., -10.){\includegraphics{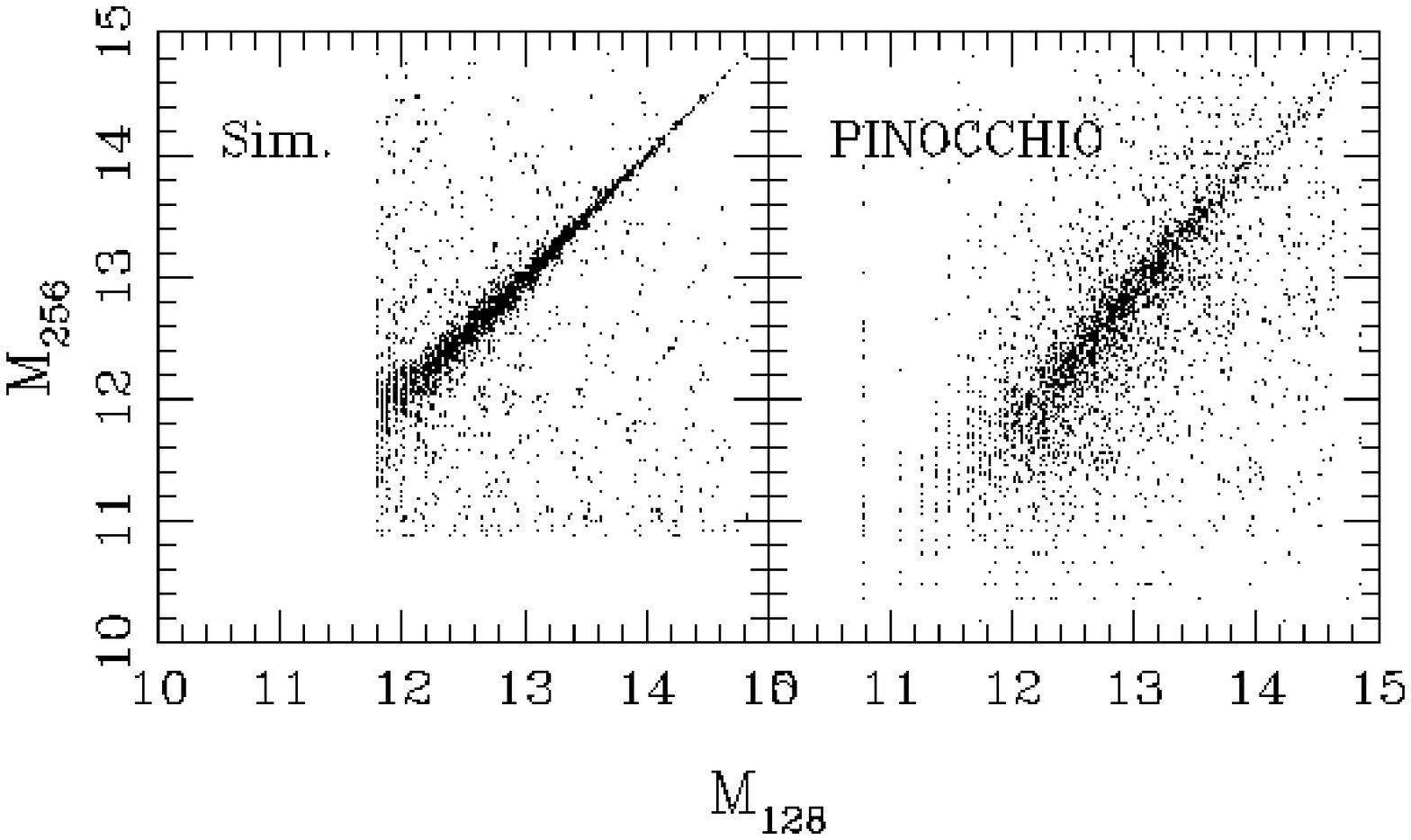}}
\end{picture}
\caption{The effects of numerical resolution for simulations and for
  PINOCCHIO. Halo masses for a random set of particles from the
  $256^3$ \LCDMd\ realisation as determined from the simulation and by
  PINOCCHIO (left and right panels respectively) are compared with the
  masses from the \LCDMt\ simulation.}
\label{fig:res}
\end{figure*}

In figure~\ref{fig:scatter} we show the accuracy with which PINOCCHIO
is able to estimate mass, Eulerian position and velocity of the
cleanly assigned objects.  In particular, we show both for SCDM and
\LCDMd\ the scatter plots of the masses, and of velocity and position
along one coordinate axis.  For comparison, the scatter plot of the
displacements of FOF halos from the initial to the final positions are
shown as well. Masses are recovered with an accuracy of $\sim$30 per
cent for SCDM and $\sim$40 per cent for \LCDMd, nearly independent of
mass. The average value is slightly biased, which results from our
constraint in reproducing the mass function.  Positions are recovered
with a 1D accuracy of $\sim$1 Mpc, slightly depending on the box size
and much smaller than the typical displacements, while velocities are
recovered with a 1D accuracy of $\sim$150 or 100 km/s for SCDM or
\LCDMd.  In general, the velocities of the fastest moving halos are
underestimated.  This could be fixed by extending the calculation of
velocities to third order LPT, although a straightforward extension
has been found not to work.

The comprehensive analysis of the statistical properties of the
PINOCCHIO halos and their merger histories, presented here and in
Taffoni et al. (2001), demonstrate that the statistical properties of
halos are always well reproduced for $N>30-50$ particles, so that the
degrading of quality of the object-by-object comparison with time
(non-linearity) is due to random noise, which does not induce
significant systematics and thus does not hamper the validity of the
halo catalogues.

We stress that these comparisons are pure predictions of PINOCCHIO, in
the sense that the free parameters of the method are constrained by the
$z=0$ mass function alone. The good agreement with the numerical
simulations confirms that PINOCCHIO is a successful approximation to the
gravitational collapse problem in a cosmological and hierarchical
context.

\subsection{Resolution effects}

As discussed above, PINOCCHIO halos resemble the FOF ones closely if
they possess a minimum number of particles of around 30-100.
Statistical quantities are well reproduced for halos with at least
30-50 particles.  These limits are comfortably similar to the minimum
number of particles needed by a {\em simulation} to produce reliable
halos.  To show this we plot in figure~\ref{fig:res}, for a random set
of particles, the mass fields (i.e. the mass of the halo the particle
belongs to) as determined by the 128$^3$ or 256$^3$ $\Lambda$CDM runs,
both for the simulations and for PINOCCHIO. The result is shown at
$z=0$. There is considerable scatter between the masses of the halos
determined from simulations with different resolutions. This scatter
is less than between PINOCCHIO and simulations, but not by much. This
result is similar at higher redshifts.  More details are given in
Appendix B, where it is shown that the match of the \LCDMd\ and
\LCDMt\ halo catalogues shows a drop in the number of cleanly
assignments for halos smaller than $\sim$30 particles
(figure~\ref{fig:ultima}a), very similar to that shown in
figure~\ref{fig:obj}.

This results suggests that resolution affects PINOCCHIO in a similar
way as it affects numerical simulations. Better resolution leads to
increased scatter in the identification of halos, since the structures
become more non-linear. For instance, we have verified that more
massive halos are reconstructed slightly {\em better} by the 128$^3$
PINOCCHIO run than by the 256$^3$ one. This is because at higher
resolution, PINOCCHIO may decide to break-up a more massive halo in
two. The degrading of the quality is modest and amounts to increased
random noise which does not bias significantly the statistics of the
halos.

\section{Angular momentum of the DM halos}

Halos are thought to acquire their angular momentum from tidal torques
exerted by the large-scale shear field while they are still in the
mildly non-linear regime (Hoyle 1949; Peebles 1969; White 1984; Barnes
\& Efstathiou 1987; Heavens \& Peacock 1988).  In this hypothesis it
is possible to estimate the angular momentum of halos using the
Zel'dovich (1970) approximation or higher-order LPT (Catelan \& Theuns
1996a,b).  The biggest difficulty in this calculation is to identify
the Lagrangian patch that is going to become a halo.

However, it was recently shown by Porciani, Hoffmann \& Dekel
(2001a,b) that the Zel'dovich approximation is unable to give very
accurate predictions of the spin of halos, as the highly non-linear
interactions of neighbouring halos tend to randomize their spins.
Assuming to know exactly which particles are going to flow into a halo
at $z=0$ and using the Zeld'ovich approximation to compute the
large-scale shear field, Porciani et al. (2001a) were able to recover
the final angular momentum of the DM halos with an average alignment
angle (defined as the angle between true and reconstructed spins) of
no better than $\sim$40$^\circ$.

Their analysis highlights the difficulty in predicting a higher-order
quantity such as the spin of DM halos.  The same calculation of spin
with N-body simulations is subject to debate.  Comparing our \LCDMd\
and \LCDMt\ simulations, we show in Appendix B that for an
order-of-magnitude estimation of angular momentum at least 100
particles per groups are required, while a more robust estimation
requires at least ten times more particles.  This is at variance with
other quantities, such as halo mass and velocity, that converge more
rapidly.  In the following we will restrict our analysis to groups
larger than 100 particles.

With respect to the analysis of Porciani et al (2001a), the PINOCCHIO
code presents the advantage of predicting with good accuracy the
instant at which particles get into the halo, while the actual shape
of the halo in the Lagrangian space is recovered with some noise,
especially in the external borders that in fact contribute most to the
angular momentum.  We have verified that the direction of the largest
axis of the inertia tensor of the halos in the Lagrangian space is
recovered within an alignment angle of $\sim$20$^\circ$, while
ellipticity and prolateness are correctly reproduced, although with
much scatter.

The estimate of the angular momentum of halos is easily performed
within the fragmentation code, with negligible impact on its speed.
When two halos with angular momenta ${\bf L}_1$ and ${\bf L}_2$ merge,
the spin ${\bf L}_{\rm merg}$ of the merger is estimated as:

\begin{equation}
{\bf L}_{\rm merg} = {\bf L}_1 + {\bf L}_2 + {\bf L}_{\rm orb},
\label{eq:spin}
\end{equation}

\noindent
where ${\bf L}_{\rm orb}$ is the orbital angular momentum of the two
halos:

\begin{equation}
{\bf L}_{\rm orb} = M_1( \Delta {\bf q}_1
                  \times \Delta {\bf v}_1 )
                  + M_2( \Delta {\bf q}_2
                  \times \Delta {\bf v}_2 ).
\label{eq:orbital}
\end{equation}

\noindent
Here $\Delta {\bf q}_i \equiv {\bf q}_i -{\bf q}_{\rm cm}$, $\Delta
{\bf v}_i \equiv {\bf v}_i -{\bf v}_{\rm cm}$, with $i=1,2$, ${\bf
  q}_{\rm cm}$ and ${\bf v}_{\rm cm}$ the position and velocity of the
centre of mass.  It is worth noticing that the use of Lagrangian
coordinates ${\bf q}$ is justified by the parallelism of displacements
and velocities.  Following Catelan \& Theuns (1997a), we stop the
linear growth of velocities not at the time of merger $t_{\rm merge}$
but at the time $t_{\rm grow}$ defined as:

\begin{equation}
t_{\rm grow} = 0.5 t_{\rm merge}
\label{eq:detach}
\end{equation}

\noindent
where $t$ is physical time.  This is a suitable generalisation of the
concept of \lq detaching\rq\ of the perturbation from the Hubble flow.
The case of accretion is treated as a merger with a 1-particle halo
which carries zero spin.

\begin{figure}
\centerline{
\psfig{figure=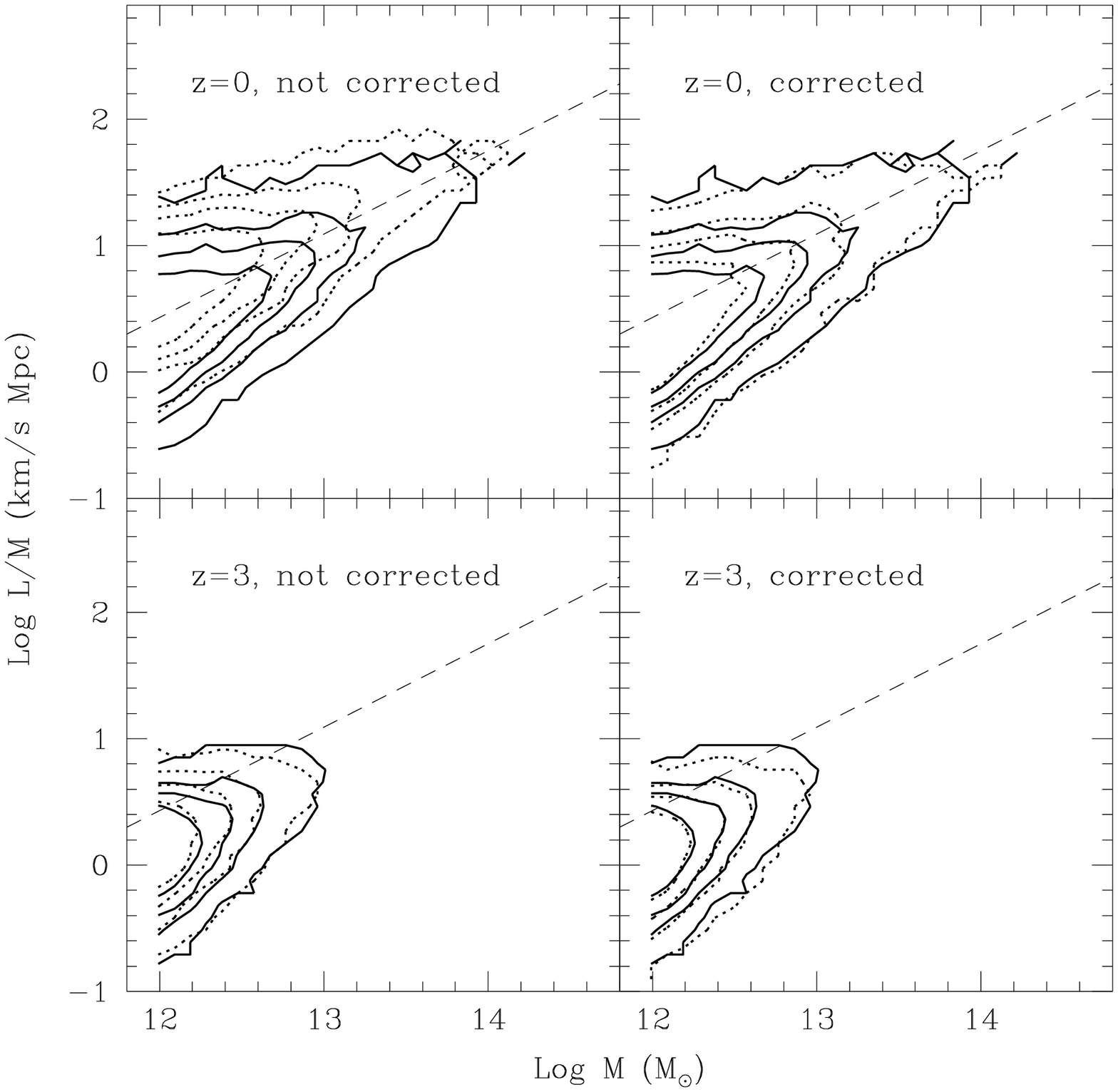,width=9cm}
}
\caption{ Mass--spin relation for dark-matter halos.  Contour lines
trace the levels of 0.2, 1, 2 and 4 halos per decade in $\log M$
($M_\odot$) and $\log L/M$ (km/s Mpc, physical units).  Continuous and
dotted lines show the contours for FOF and PINOCCHIO halos
respectively.  Dashed lines give the scaling $L\propto M^{5/3}$ (Catelan
\& Theuns 1996a).  Upper panels: $z=0$; lower panels: $z=3$.  Left
panels: no correction; right panels: spin corrected as in
equation~\ref{eq:shuffle}.  }
\label{fig:spinmass}
\end{figure}

The so-obtained angular momenta obey a mass--spin relation which is
roughly consistent with that of the FOF groups.  This is shown in the
left panels of figure~\ref{fig:spinmass} for the \LCDMd\ simulation.
Although qualitatively similar, the PINOCCHIO relation overestimates
the FOF one by some factor which is larger for the smaller halos.  If
the lower value of the spin is due to the higher degree of non-linear
shuffling suffered by halos because of tidal interaction with
neighbours, this trend of having lower-mass halos more randomized than
higher-mass ones is in agreement with that suggested by Porciani et
al. (2001a).

It is useful to improve this prediction, so as to obtain angular
momenta for the halos with accurate statistical properties.  To this
aim we decrease each component of the spin at random, following the
simple rule:

\begin{equation} L_i^{\rm new} = L_i \times ((1-f_{\rm spin}) + f_{\rm spin} 
\times f_{\rm rand}),
\label{eq:shuffle}
\end{equation}

\noindent
where $f_{\rm spin} = f_0 + f_1 (M/M_*(z))$ (forced to $0\le f_{\rm
  spin} \le 1$) and $f_{\rm rand}$ is a random number ($0<f_{\rm
  rand}<1$).  The two parameters $f_0$ and $f_1$ are fixed so as to
reproduce at best the mass--spin relation of
figure~\ref{fig:spinmass}.  Optimal values are $f_0=0.8$ and
$f_1=0.15$.  The right panels of figure~\ref{fig:spinmass} show the
resulting mass--spin relations, which agrees fairly well with the FOF
ones.

\begin{figure}
\centerline{
\psfig{figure=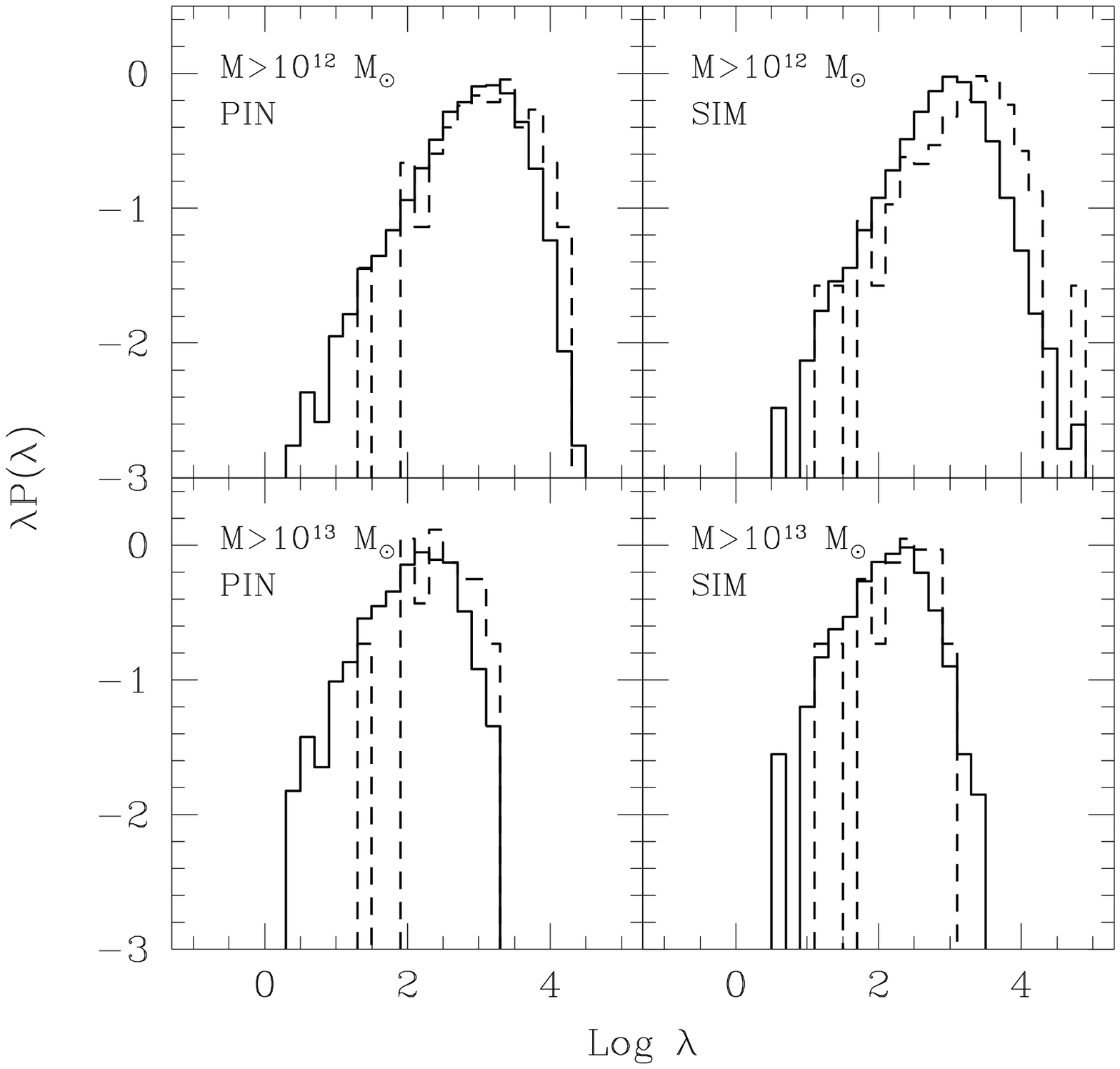,width=9cm}
}
\caption{
  Correlation between spin and merging history.  Continuous lines: all
  halos; dashed lines: halos that have suffered a major merger.  Left
  panels: PINOCCHIO predictions; right panels: simulation.  Upper
  panels: halos with $M\ge 10^{12}\ M_\odot$; lower panels: halos with
  $M\ge 10^{13}\ M_\odot$.}
\label{fig:spinhist}
\end{figure}

Apart from the mass--spin correlation shown in
figure~\ref{fig:spinmass}, the angular momentum is known to be nearly
independent of other halo properties (Ueda
et al. 1994; Cole \& Lacey 1996; Nagashima \& Gouda 1998; Lemson \&
Kauffman 1999; Bullock et al. 2001; Gardner 2001; Antonuccio-Delogu et
al. 2001), with the exception of a weak dependence with the merger
history of the halos.  The dependence of spin on the environment is
still debated (Lemson \& Kauffman 1999; Antonuccio-Delogu et
al. 2001).  Gardner (2001) has shown that halos that have suffered a
major merger tend to have higher spin.  In figure~\ref{fig:spinhist}
we show that this trend is successfully reproduced by PINOCCHIO halos.
Merged halos at $z=0$ have been selected by requiring that the second
largest progenitor halo at $z=0.25$ is larger than 0.3 times the final
halo mass.  To extract the mass--spin relation, we define the quantity
$\lambda \equiv Log L - 1.5 (Log M/M_*)$.  As apparent in
figure~\ref{fig:spinhist}, the $\lambda$-distribution of the merged
halos is biased toward larger $\lambda$-values both for the simulation
and for the PINOCCHIO halos, although the trend may be slightly
underestimated by PINOCCHIO.

\begin{figure}
\centerline{
\psfig{figure=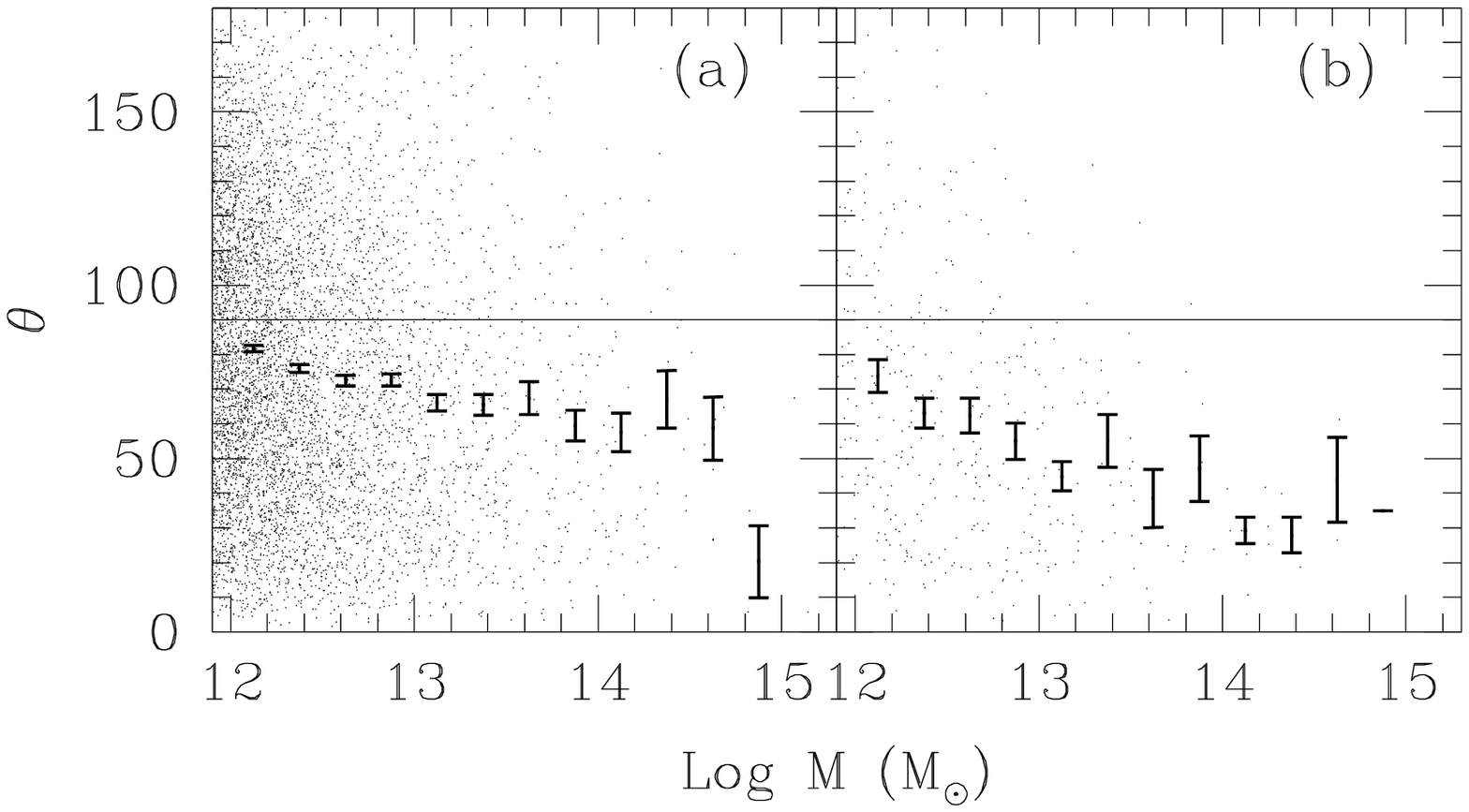,width=9cm}
}
\caption{ Alignment angle $\theta$ for pairs of cleanly assigned halos
at $z=0$ as a function of mass.  The errorbars denote averages in mass
bins, errorbars give the rms of the mean.  (a) all halos, (b) halos
with $f_{\rm ov}>0.7$.}
\label{fig:spinobj}
\end{figure}

The agreement at the object-by-object level is in line with the
intrinsic limits of perturbative theories found by Porciani et al.
(2001a).  Figure~\ref{fig:spinobj} shows the alignment angle $\theta$
for the spins of cleanly matched FOF and PINOCCHIO halos, and their
average values computed in bins of mass (error flags indicate the rms
around the mean).  While the left panel shows all halos, the right
panel is restricted to those pairs of halos that overlap by more than
70 per cent.  The average angle is significantly smaller than
90$^\circ$, highlighting a significant correlation of PINOCCHIO and
FOF spins.  However, the alignment is at best as high as
$\sim$60$^\circ$.  This is mostly due to errors in the definition of
the halo, as shown by the right panel, where the best reconstructed
halos with more than 1000 particles show an average alignment angle of
$\sim$30--40$^\circ$, consistent with the intrinsic limit quoted by
Porciani et al. (2001a).

To conclude, the prediction of angular momentum of halos is severely
hampered by the intrinsic limits of linear theory described by
Porciani et al. (2001a) and further worsened by the error made by
PINOCCHIO in assigning particles to halos.  The correct statistics is
reproduced only by introducing two more \lq fudge\rq\ free parameters,
while the object-by-object agreement is poor although significant.
However, even N-body simulations do not converge rapidly in
estimating this quantity (see Appendix B).  Moreover, the important
spin--merger correlation is recovered naturally.  Although we do not
claim this result as a big success, we notice that PINOCCHIO is, to
our knowledge, the only perturbative algorithm able to predict the
spin of halos at the object-by-object level.  Moreover, the prediction
of spins comes at almost no additional computational cost, and the
whole acquisition history of angular momentum can be followed for each
halo.  Thus, we regard the use of the angular momenta provided by
PINOCCHIO as a viable alternative to drawing them at random from some
distribution that fits N-body simulations (Cole et at. 2000;
Vitvitsaka et al. 2001; Maller, Dekel \& Somerville 2001).

\section{Discussion}

PINOCCHIO is an approximation to the full non-linear gravitational
problem of hierarchical structure formation in a cosmological setting,
in contrast to the mostly statistical approaches such as the PS
prescription. The good agreement in detail between PINOCCHIO and FOF
halos identified in simulations, explains the ability of the method to
generate reliable halo catalogues. It also demonstrates that the
underlying dynamical approximations work well. With respect to the
results of Monaco (1995; 1997a,b), PINOCCHIO addresses successfully
the geometrical problem of the fragmentation of the collapsed medium
into objects and filaments.

While a direct analytical rendering of the fragmentation prescription
as used in PINOCCHIO seems very complex, because it requires knowledge
of spatial correlations to high order, analytical progress might
nevertheless be possible. For instance, Monaco \& Murante (1998)
proposed to generalise the mass-radius relation of PS, to allow a more
general distribution of masses to form at a given smoothing
radius. This was formulated in terms of a \lq growing\rq~ curve for the
objects, that gives the fraction of mass acquired by the object at a
given smoothing radius. The mass function is then obtained by a
deconvolution of the $\Omega(<\sigma^2)$ function (as obtained from ELL
collapse, like in figure~\ref{fig:omega_rc}) with the growing curve of
the objects.  This growing curve could be estimated from the results of
PINOCCHIO, giving an improved analytical expression for the mass
function.  But in the case of Gaussian smoothing merging histories
cannot be computed from the excursion set formalism, because the
trajectories are strongly correlated (Peacock \& Heavens 1990; Bond et
al. 1991), so that the random walk formalism cannot be used.  Moreover,
it is impossible from such an approach to have full information on the
spatial distribution of objects.  So, such analytic extensions of
PINOCCHIO would not be as powerful as the full analysis.  Besides,
analytic formalisms based on peaks (Manrique \& Salvador Sole 1995;
Hanami 1999) are manageable only when linear theory is used.  We
therefore regard methods like PINOCCHIO which are based on an actual
realisation of the linear density field, as a good compromise between
performing a simulation, and getting only statistical information from
a PS like approximation.

As mentioned in the introduction, similar methods have been proposed
in the literature, such as the peak-patch method of Bond \& Myers
(1996a), the block model of Cole \& Kaiser (1988), and the merging
cell model of Rodrigues \& Thomas (1996) and Lanzoni et al. (2000).  A
qualitative comparison with peak-patch reveals a similar accuracy in
reproducing the masses of the objects. From figure 10 of Bond \& Myers
(1996b) it is apparent that, in a context analogous to our SCDM
simulation, masses are recovered with an accuracy of $\sim$0.2 dex,
not much worse than the one given in our figure~\ref{fig:scatter} for
SCDM.  Unfortunately, it is not clear from the Bond \& Myers papers to
which extent the agreement can be pushed down to galactic masses.  As
linear theory under predicts the fraction of collapsed mass when the
variance is large, a deficit of peaks corresponding to smaller masses
is possible.  The objects selected by peak-patch are constrained to be
spherical in the Lagrangian space (they collapse like ellipsoids but
start-off as spheres perturbed by the tidal field), while PINOCCHIO is
not restricted in this sense and is able to reproduce the orientation
of the objects in the Lagrangian space, as mentioned in Section 5.
Moreover, PINOCCHIO is not affected by the problem of peaks
overlapping in the Lagrangian space.  Finally, peak-patch has never
been extended, to the best of our knowledge, to predict the merger
histories of objects.

The merging cell model of Lanzoni et al. (2000) shares some properties
with PINOCCHIO, in particular the fact that both codes build-up halos
through mergers and accretion. However, the non-linear ellipsoidal
collapse of PINOCCHIO is an important improvement, as is the use of
Gaussian filters instead of box car smoothing. Also, the size of the
merging objects tend to be quite large in the merging cell model,
whereas PINOCCHIO allows accretion of single particles. We have been
able to compare our results directly with those of Lanzoni et al.
(2000). The halos identified in the merging cell model do not
accurately reproduce those from the simulations.  This poorer level of
agreement is partly due to the cubic shape of the cells and to the
coarse resolution of the box car smoothing.  As a consequences of
these choices, massive halos appear as big square boxes, and the mass
function shows fluctuations with spacing of factors of two that
reflect the smoothing.

\subsection{First-axis versus third-axis collapse}

Recently there has been extensive discussion in the literature about
whether the collapse of the first axis is enough to characterise
gravitational collapse, or whether all three axes should reach
vanishing size (Bond \& Myers 1996a; Audit et al. 1997; Lee \&
Shandarin 1999; Sheth et al. 2001).  Here we try to clarify this issue,
showing that apparently contradictory claims result from different
interpretations of ellipsoidal collapse, and from the choice of
smoothing window.

As described in section 2.1, ellipsoidal collapse can be considered as
a truncation of LPT, a convenient description of the dynamical
evolution of a mass element. In other words, ELL does not attempt to
describe the collapse of an {\em extended} ellipsoidal peak, rather,
it operates on the infinitesimal level. Given this, OC appears as the
most sensible choice for the collapse condition, for the reasons
already outlined in Section 2.1, and with the {\it caveat} that the
mass undergoing OC may end up either in halos or in filaments.  OC
corresponds to collapse along the first axis, which means that the
ellipsoid has undergone pancake collapse.  However, this does not
imply that the extended region is flattened as well.  Indeed, as the
example in Monaco (1998) illustrates, in the collapse of a spherical
peak with decreasing density profile, all mass elements (except for
the one in the centre) collapse as needles pointing to the centre.
This is because the spherical symmetry guarantees that the first and
second axis collapse together. Yet the collapse of the peak is {\em
  not} that of a filament but of a sphere. This shows how misleading
the local geometry of collapse is for understanding the global
geometrical properties of the collapsing matter.

Alternatively, ellipsoidal collapse can be used to model extended
regions associated to a particular set of points, such as density
peaks (Bond \& Myers 1996; Sheth et al. 2001).  In this case,
first-axis collapse truly corresponds to the formation of a flattened
structure, while third-axis collapse corresponds to the formation of a
spheroidal object.  For instance, in the case of the spherical peak
mentioned above, the peak point is collapsing in a spherical way both
locally and globally.  It is clear that in such cases a satisfactory
definition of collapse must be related to third-axis collapse.  Sheth
et al. (2001) showed that indeed this collapse condition improves the
agreement with simulations when the centres of mass of FOF objects are
considered (a particular set of points analogous to the peaks), but
does not help much when general unconstrained points are considered.

The two definitions of collapse are very different from many points of
view.  First-axis collapse is on average faster than the spherical one
(Bertschinger \& Jain 1994), while third-axis collapse is
correspondingly slower.  Moreover, while 50 per cent of mass is
predicted to collapse at very late times by linear theory (starting
from a density field with finite variance and not taking into account
the cloud-in-cloud problem), $23/25\sim $ 92 per cent of mass is
predicted to undergo first-axis collapse, but only 8 per cent
third-axis collapse.  This is very important when computing the mass
function with a PS-like approach: while first-axis collapse more or
less reproduces the correct normalisation (Monaco 1997b), third-axis
collapse requires a large \lq fudge factor\rq\ $\sim $12 (Lee \&
Shandarin 1998), as only 8 per cent of mass is available for collapse.

Whithin the framework of the excursion set approach, it is interesting
to understand whether the introduction of ellipsoidal collapse is
going to improve the statistical agreement between simulations and PS.
Monaco (1997b) and Sheth et al. (2001) showed that ellipsoidal
collapse can be introduced through a \lq moving\rq\ barrier which
depends on the variance $\sigma^2$ of the smoothed field.  Third-axis
collapse gives longer collapse times than spherical collapse, and this
corresponds to a barrier which rises with $\sigma^2$, while the
opposite is true for first-axis collapse.  In the case of {\em sharp
  $k$-space smoothing}, the fixed barrier reproduces the PS mass
function and hence overestimates the number of low mass objects. Sheth
et al.~(2001) showed that using the moving barrier appropriate for
third-axis collapse leads to the formation of fewer low mass objects,
and hence improves the mass function. However, when {\em Gaussian
  smoothing} is used, the fixed-barrier solution is different from PS,
and the number of small mass halos is now severely {\em
  underestimated}.  Monaco (1997b, 1998b) showed that in this case
first-axis collapse (with no free parameter to tune) produces a
reasonable fit to the simulations, with some improvement with respect
to PS.

From these considerations, it is clear that a successful definition of
collapse depends on many technical details, such as the kind of
dynamics considered (mass elements versus extended regions) and the
type of smoothing used (sharp $k$-space versus Gaussian smoothing). We
choose to consider Gaussian smoothing and first-axis collapse (OC)
applied to mass elements.  These choices are consistent with the
excursion set approach, but need to be supplemented by an algorithm to
fragment the collapsed medium into halos and filaments. This is because
the collapse definition operates on mass elements and does not specify
the larger structures that collapse together. Moreover, the strong
correlation of Gaussian trajectories in the $F-R$ plane implies that
merger histories cannot be recovered with the same simple and elegant
algorithm used by Bond et al. (1991) and Lacey \& Cole (1993).  The
alternative algorithm proposed by Sheth et al. (2001), based on sharp
$k$-space smoothing, has the advantadge of being analytical and simpler
than PINOCCHIO.  However, the choice of third-axis collapse is
physically motivated by comparing the collapse of the centres of mass
of FOF halos to the simulations, but the probability distribution used
in the excursion set approach is the unconstrained one of the general
points.  Moreover the mass of the objects is still estimated as if
top-hat smoothing were used.  So, we regard Sheth et al. (2001) as
another phenomenological model, yet improved with respect to PS and
very effective in providing statistical information.

\section{Conclusions}

We have presented a detailed description of PINOCCHIO, a fast and
perturbative approach for generating catalogues of DM halos in
hierarchical cosmologies.  Given a set of initial conditions,
PINOCCHIO produces masses, positions, velocities and angular momenta
for a catalogue of halos.  Because PINOCCHIO is based on
reconstructing the mergers of halos, accurate information on the
progenitors of halos is available automatically (Taffoni et al. 2001).
We have compared in detail these catalogues with two N-body
simulations which use different cosmologies, resolutions, mass ranges
and N-body codes.  The match is very good, both for statistical
quantities, which are recovered with a $\sim$5 per cent accuracy for
the mass function and $\sim$20 per cent for the correlation function
($\sim$10 per cent error in $r_0$), and at the object-by-object level,
whereas a $\sim$20-40 per cent accuracy in halo mass is reached for
$>$70-100 per cent of the objects that have at least 30--100
particles.  These results show that PINOCCHIO is a proper
approximation of the gravitational problem, and not simply a
phenomenological model able to reproduce some particular aspects of
gravitational collapse.

PINOCCHIO consists of two steps.  In the first step, the estimate of
collapse time, no free parameter is introduced, as collapse is defined
as OC.  The second step addresses the geometrical problem of the
fragmentation of the collapsed medium into objects and the
disentanglement of the filament web.  This is analogous to the process
of clump finding in N-body simulations, and requires the introduction
of free parameters, at least one to specify the level of overdensity
at which halos are selected (indeed, while algorithms like FOF or SO
introduce only one parameter, others like HOP introduce more than
one).  In fact five free parameters (that are not independent) are
introduced, two to characterize the events of merging and accretion,
and the others for fixing resolution effects.  They are tuned by
reproducing the FOF numerical mass function with linking length equal
to 0.2.

PINOCCHIO is fast and can be run even with small computers: all the
tests presented in this paper were run with a simple PC with Pentium
III 450MHz processor and 512M of RAM.  For a grid of 256$^3$ particles
the first step runs in $\sim$6 hours independent of the degree of
non-linearity, while the second step requires only a few minutes.  The
typical outcome of such a run is a catalogue of many thousands of
objects with known positions, merger histories and angular momenta.
With a supercomputer one could run tens of large, say 512$^3$,
realisations in a fraction of the time required by a single one to be
run with a standard N-body code, and obtain all the merger histories
without the expensive post-processing analysis required in the case of
simulations.

The results of PINOCCHIO are suitable for studies of astrophysical
events in a cosmological context, as they give essentially most of the
information that a large-volume N-body simulation can give.  In
particular, the availability of catalogues with final positions,
merger histories and angular momentum makes PINOCCHIO a suitable tool
to be used in the context of galay formation.  A public version of
PINOCCHIO is available at the web site
http://www.daut.univ.trieste.it/pinocchio.

\section*{Acknowledgments}
The authors thank Stefano Borgani, Fabio Governato, Barbara Lanzoni
amd Cristiano Porciani for many discussions.  Fabio Governato, Tom
Quinn and Joachim Stadel have kindly provided the 360$^3$ SCDM
simulations used in this paper.  PM acknowledges support from MURST.
TT acknowledges support by the \lq Formation and Evolution of
Galaxies\rq~ network set up by the European Commission under contract
ERB FMRX-CT96086 of its TMR programme, and thanks PPARC for the award
of a post-doctoral fellowship.  Research conducted in cooperation with
Silicon Graphics/Cray Research utilising the Origin 2000 super
computer at DAMTP, Cambridge.

{}

\appendix
\section{parameters for merging and accretion}

The accretion and merging conditions given in equations~\ref{eq:accr}
and \ref{eq:merg} work well when the halos contain sufficiently many
particles.  However, for smaller halos, the limiting distance $f_{\rm
a} R_{\rm N}$ or $f_{\rm m} R_{\rm N}$ may be comparable to the grid
spacing.  In this case, the Zel'dovich velocity \vmax\ needs to be very
accurate in order that accretion or merging to take place, and this may
lead PINOCCHIO to underestimate the number of very low mass
objects. The simplest solution to this problem is to add a constant
$f_{\rm r}$ to the right hand side of equations~\ref{eq:accr} and
\ref{eq:merg}, of order of the grid spacing, as was done in
paper~I. This brings the number of parameters in PINOCCHIO to three,
$f_{\rm a}=0.18$, $f_{\rm m}=0.35$ and $f_{\rm r}=0.70$.

However, when applied to the \LCDMd\ simulation, this choice produces
a systematic excess of low mass objects at high redshift, of order
$\sim$20 per cent at $z=4$ for objects of 30 particles.  This excess
is barely noticeable in the SCDM simulation at $z=1.13$ (see figure 1
of paper I). The origin of this systematic effect is the following.
The accuracy of LPT in estimating the velocities is not constant in
time, but depends on the degree of non-linearity reached, worsening at
later times.  It can be measured by comparing the Zel'dovich
displacements with those from the simulation, for particles that are
just experiencing OC collapse, according to the \fmax\ field.

\begin{figure}
\centerline{
\psfig{figure=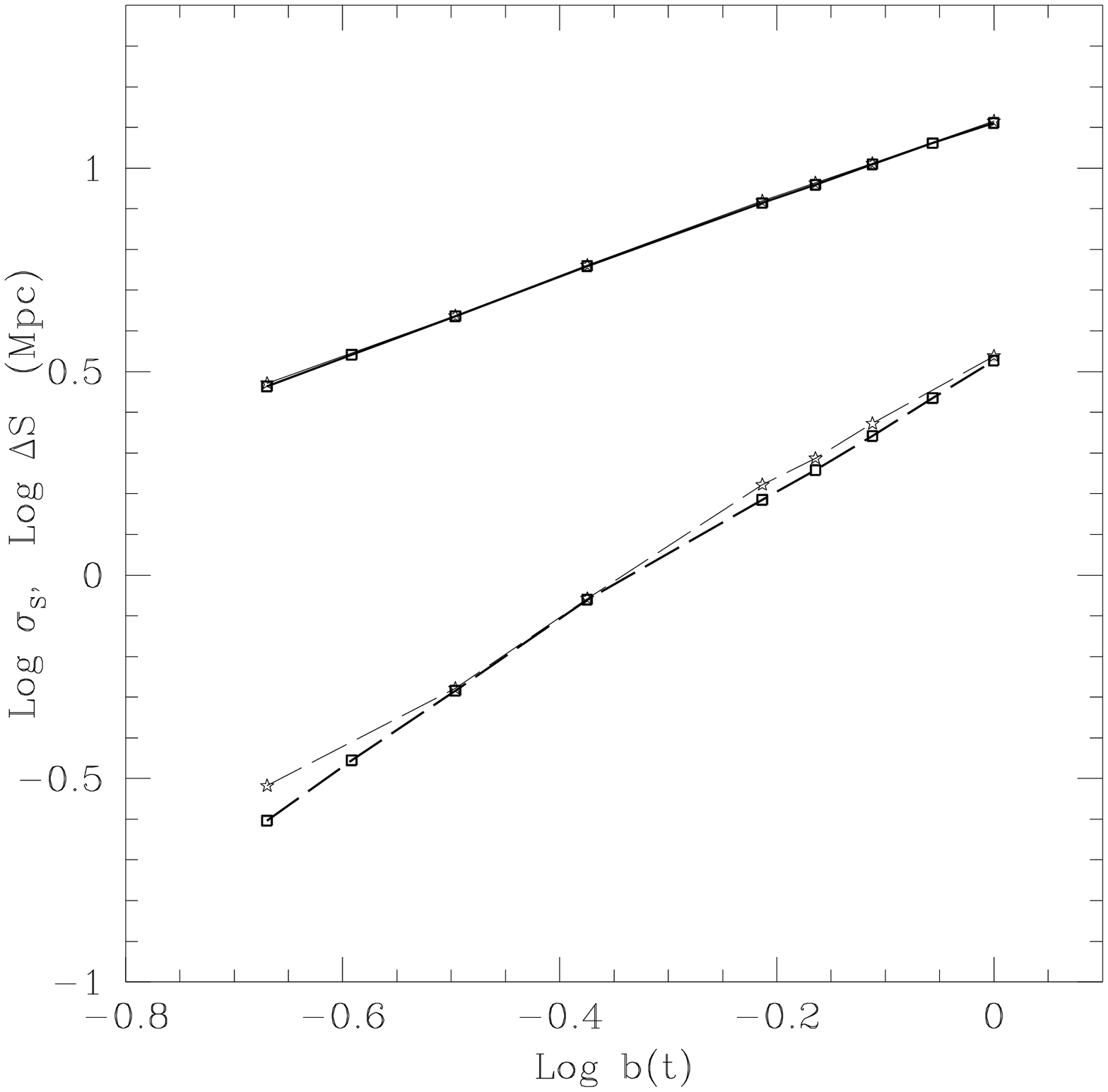,width=9cm}}
\caption{Error in the estimate of the Zel'dovich displacements for
  particles that have just undergone orbit crossing, as function of
  the growing mode $b$.  Continuous lines are the average displacement
  of the collapsing particles, dashed lines are the error in the
  estimate of these displacements, as computed by comparing with the
  simulation results.  Thick lines are obtained from the \LCDMd\ 
  simulation, thin lines from the \LCDMt\ simulation.}
\label{fig:error}
\end{figure}

In figure~\ref{fig:error} we show that the error in the displacement
increases as the field becomes more non-linear. The rate of increase is
very similar for the two cosmological models plotted. The errors in the
displacements are much smaller than the displacements themselves,
demonstarting the power of the Zel'dovich approximation.  While the
average displacement grows as $b(t)$, its error grows as $b(t)^{1.7}$.

The fact that displacements are computed more accurately at earlier
times has two important consequences. Firstly, the accuracy of the
reconstruction of particle position will degrade with time, as we
illustrated in Section 4.  Secondly, objects at higher redshifts will
tend to accrete mass more easily than at later times, for a given set
of parameters $f_{\rm a}$, $f_{\rm m}$ and $f_{\rm r}$. The reason is
that, if a particle should accrete onto a halo at late times, we need
to make these parameters sufficiently genereous so that the particle
falls within $d$ of the halo according to Eq.~\ref{eq:accr}, even
though we are unable to compute the position of the particle very
accurately. But as a result, this may lead to too much accretion at
earlier times, when the positions are more accurate.

It is possible to improve PINOCCHIO to correct for this numerical
problem.  What is relevant in the fragmentation code is not the
absolute displacement of a particle, but the displacement relative to
that of the halo. The distance between a collapsing particle and the
centre of mass of a group is $d \sim S_{a,b}\times R_{\rm N}$.
Considering that $S_{a,b}\propto b$, its variance scales as the
variance $\sigma^2$ of the linear density and the relative error on
$S_{a,b}$ grows $\propto b(t)^{0.7}$, we can estimate the uncertainty
on $d$, given the errors in reconstructing positions as

\be \delta d = f_{\rm s} \sigma(R_{\rm N}) R_{\rm N} b(t)
[\sigma(R_{\rm N}) b(t)]^{0.7}. \label{eq:fs} \ee

\noindent
Here, $f_{\rm s}$ is another free parameter. We only introduce this
extra parameter in the accretion condition, since the results do not
improve when we apply a similar correctio to the merger condition.  The
accretion and merging conditions are then:

\be d < f_{\rm a} \times R_{\rm N} + f_{ra} + \delta d
\label{eq:newaccr} \ee

\be d < f_{\rm m} \times \max (R_{\rm N1},R_{\rm N2}) + f_{rm}. 
\label{eq:mergnew} \ee

\noindent
We note that the resolution parameter $f_{\rm r}$ is now different for
accretion and for merging.

Our algorithm now contains five parameters. The best fit values have
been determined by generating many realisations of Gaussian fields
(including the initial conditions of the SCDM, \LCDMd\ and \LCDMt\ 
simulations used here) for different cosmological models, box sizes
and resolutions, and determining for each realisation those parameters
that best fit corresponding mass function. We used the analytical mass
function of Jenkins et al. (2001) as template. The best fit is easily
achieved, as the effects of small variations of only one parameter are
rather simple.  In particular, $f_{\rm m}$ determines the overall
slope of the mass function, $f_{\rm rm}$ the slope at low masses,
$f_{\rm a}$ the normalisation, $f_{\rm ra}$ the abundance of low mass
halos and $f_{\rm s}$ the abundance of low mass halos at low
redshifts.

The best fit values are $f_{\rm m}=0.35$ and $f_{\rm rm}=0.7$, as in
paper~I.  The parameters for accretion are found to be correlated,

\be f_{\rm ra} = 0.40 - 3.5\,(f_{\rm a} - 0.22)\,. \label{eq:deg} \ee

\noindent
In addition, $f_{\rm ra}$ correlates with the degree of non-linearity,
as quantified by $\Sigma\equiv \sigma(R=0)/l_{\rm grid}$. Here,
$\sigma(R=0)$ is the variance at the level of the grid and $l_{\rm
grid}$ the grid spacing. $\Sigma$ is sensitive to both
the degree of non-linearity reached and the level of accuracy 
of the Zel'dovich displacements. The best fit for $f_{\rm a}$ is

\be f_{\rm a}  = 0.22 + (\log \Sigma - 0.36) * 0.11\,.
\label{eq:res} \ee

\noindent
We also demand that $0.22\le f_{\rm a}\le 0.26$. The best fit $f_{\rm
  s}=0.06$.  (In all cases a change in the last significant digit
gives differences in the mass function appreciable at the 5 per cent
level).

Unfortunately, these parameters are sensitive to the algorithm used to
generate initial conditions, in particular they depend on how the small
scale power close to the Nyquist frequency is quenched. For the \LCDMd\
models, we used the initial conditions generator distributed with HYDRA
(Couchman et al. 1995), where power below the Nyquist frequency (on a
grid with unit grid spacing, taken to be $k_e=0.8\pi$), is quenched
exponentially $\propto \exp(-(k/k_e)^{16})$. The parameters for
PINOCCHIO apply for this type of initial conditions generator.  In
contrast, the initial conditions for the 360$^3$ SCDM simulation were
generated on a 180$^3$ grid, without an additional cut-off of small
scale power. The corresponding PINOCCHIO parameters are $f_{\rm
a}=0.19$, $f_{\rm ra}=0.60$ and $f_{\rm s}=0.04$.

It is possible there are other degeneracies amongst these
parameters. We have verified through extensive analysis that the
object-by-object agreement is rather insensitive to the precise values,
once the mass function fits well, the object-by-object agreement is
good too. We have tried many other recipes for the parameters, but
this one is adequate for generating reliable halo catalogues for a wide
variety of cosmological models.

\section{Reliable estimate of the angular momentum of a DM halo}

It was mentioned in Section 5 that a reliable estimate of the angular
momentum of an N-body halo requires at least 100 particles.  As this
matter is of interest to many N-body simulators, we give here details
of the analysis we have performed.

This matter can be addressed by using our \LCDMd\ and \LCDMt\
simulations, recalling that \LCDMt\ is run on the same initial
conditions as \LCDMd, resampled on the coarser grid.  We consider the
$z=0$ outputs of the two simulations and match the halo catalogues in
exactly the same way as done in the object-by-object comparison of
PINOCCHIO and N-body catalogues.  In practice, the 256$^3$ linking
list is resampled to 128$^3$ by nearest grid assignment, i.e. simply
by considering 1 particle over 8 and skipping the others.  Notice that
this resampling is used only to match halo pairs, the halo properties
are computed from the complete lists of particles.  In the following
we will assume the properties of the 256$^3$ groups as {\it bona fide}
estimate, and will interpret the difference between 128$^3$ and
256$^3$ as the error on the lower-resolution groups.

\begin{figure*}
\centerline{
\psfig{figure=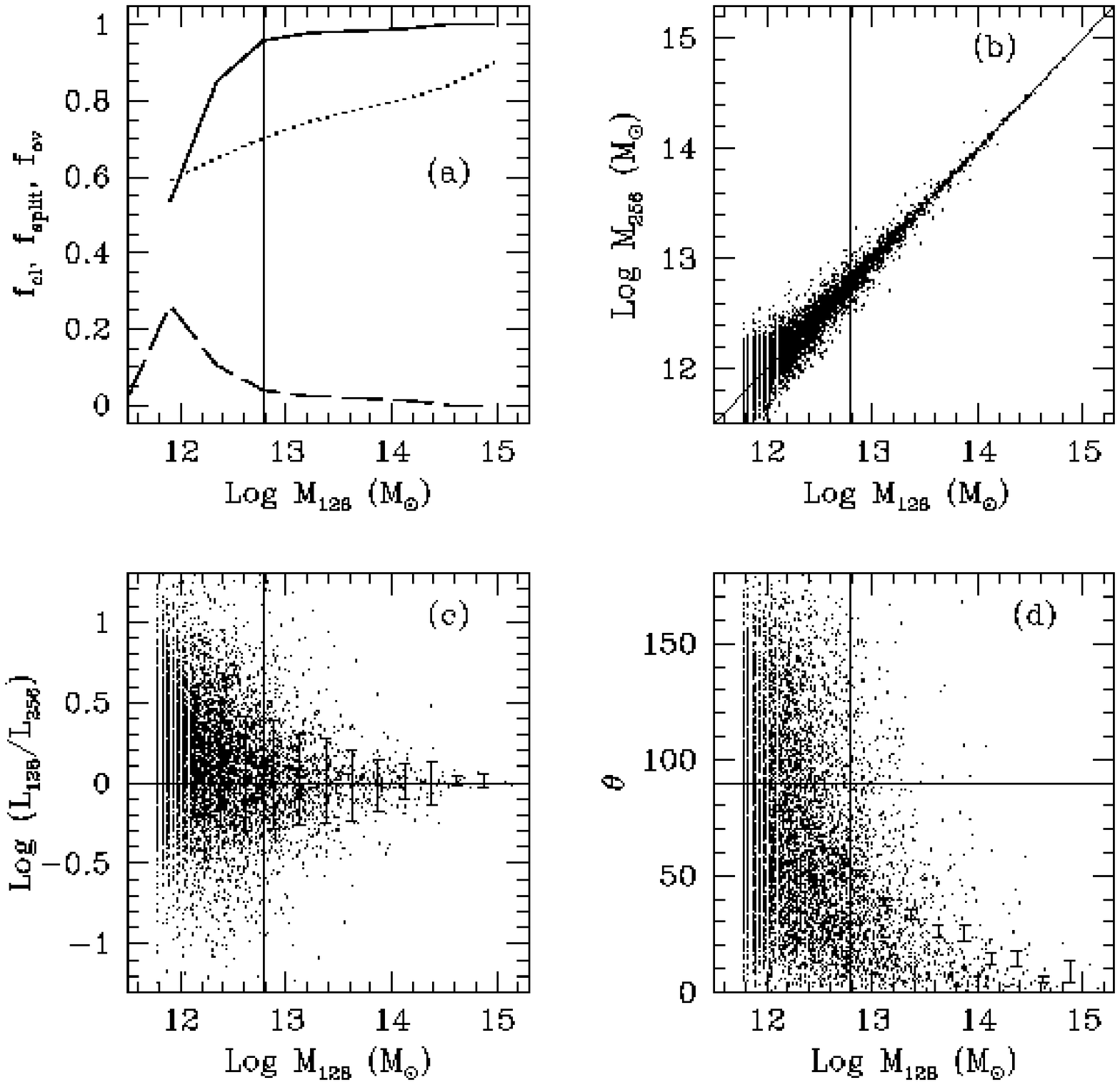,width=15cm}}
\caption{In all the panels the vertical line marks groups of 100
particles in the \LCDMt\ simulation.  (a) Matching of the \LCDMd\ and
\LCDMt\ halo catalogues.  Continuous line: \fcl; dashed line: \funcl;
dotted line: \fov.  (b) Correlation of masses for the cleanly assigned
objects.(c) Fractional difference of angular momenta for the cleanly
assigned objects, as a function of mass.  Error bars give the rms
difference in bins of mass.  (d) Alignment angle between the angualr
momenta of cleanly assigned objects.  Error bars give the rms of the
mean of the alignment angles in bins of mass.}
\label{fig:ultima}
\end{figure*}

Figure~\ref{fig:ultima}a shows the fractions \fcl, \funcl\ and \fov\
for the matching of the two catalogues, as a function of the mass of
the halo according to the \LCDMt\ simulation.  In this and subsequent
panels the vertical line marks the groups of 100 particles (128$^3$).
The matching of the two catalogues is excellent for groups larger than
100 particles, but still reasonable for groups as small as $\sim$30
particles.  Mass estimates are pretty stable
(figure~\ref{fig:ultima}b), with an error of 30-40 per cent for the
smallest groups, decreasing to the high-mass end.

Conversely, the error on the spin estimate turns out to be much
larger.  Figure~\ref{fig:ultima}c shows the fractional difference
between halo spins as a function of mass (the rms difference is also
shown), while figure~\ref{fig:ultima}d shows the alignment angles of
the spins (the rms of the mean is shown in this case, as in
figure~\ref{fig:spinobj}).  The rms difference is still in excess of a
a factor of two for halos of 100 particles, and even larger for
smaller halos.  Moreover, the spin directions of small halos are very
poorly correlated for halos with less than 100 particles.  We conclude
that the lower limit for a correct order-of-magnitude estimate of the
angular momentum of a halo is 100 particles, while a more precise
estimate will require at least ten times more particles.

\end{document}